\documentclass[a4paper,11pt]{article}
\pdfoutput=1 

\usepackage{jcappub} 

\usepackage[T1]{fontenc} 

\title{How the cosmic voids contribute to stalling and quenching the giant galaxies on their surfaces}

\author{Geonwoo Kang,}
\author[1]{Jounghun Lee \note{Corresponding author.}}

\affiliation{Department of Physics and Astronomy, Seoul National University, \\
Kwanak-ro 1, Kwanak-gu, Seoul 08826, Republic of Korea}

\emailAdd{kanggeonwoo@snu.ac.kr}
\emailAdd{cosmos.hun@gmail.com}

\abstract{We report a numerical hint that the formations of cosmic voids may be closely linked with the mechanism through which the giant galaxies on void surfaces establish 
elliptical shapes, redder colors, and lower specific star formation rates (sSFR). Identifying the voids from the TNG300-1 simulations via the Void-Finder 
algorithm~\cite{HV02} at $z=0$, $0.5$ and $1$, we explore if and how the shapes of the TNG galaxies located on void surfaces are aligned with the directions toward the void centers.  
Noting that only the giant void-surface galaxies with stellar masses $M_{\star}\ge 10^{10.5}\,h^{-1}\,M_{\odot}$ exhibit significant tendency of perpendicular alignments, 
we dichotomize them into two $M_{\star}$-controlled samples according to their morphologies (elliptical or spiral), colors (redder or bluer), sSFR (lower or higher) and stellar ages 
(older or younger).   It is found at all of the three redshifts that the perpendicular alignments of void-surface galaxies become stronger for the cases that they have elliptical shapes, 
redder colors, and lower sSFR,  but showing weak dependence on the stellar ages. It is also shown that the numerical results are well described by the analytical one-parameter model developed 
by Lee~\cite{lee19} under the assumption of the existence of a linear scaling between the covariance matrices of galaxy shape axes and local tidal tensors. 
We test the robustness of alignment signals against the variation of void-finder algorithms and its feasibility against the redshift-space and projection effects. 
Our results lead us to speculate that the formation and expansion of voids may have an effect of stalling and quenching the giant void-surface galaxies by compressing adjacent matter and 
then preventing them from radial infall/accretion.}
\begin{document} 
\maketitle
\flushbottom

\section{Introduction}\label{sec:intro}

The large-scale spatial distributions of galaxies in the universe display anisotropic network-like fabric, which is collectively called the cosmic web~\cite{web96}. 
The presence of the cosmic web is well fitted into the standard picture of hierarchical structure formation based on the gravitational instability theory~\cite{zel70}, 
according to which the main mechanism is nothing but the occurrence of anisotropic collapse of matter along three distinct principal axes of the large-scale 
tidal fields~\cite{zel70,khl-etal85,bbks86,web96,cau-etal14}. 
The large-scale structures that constitute the cosmic web are categorized into four distinct types, namely, voids, sheets, filaments and knots, according to their 
structural dimensions~\citep{zel70,ara-etal07,hah-etal07}. 
Most of the galaxies are embedded in one-dimensional filaments~\cite{ara-etal10,lib-etal18,gan-etal19} formed through the gravitational collapse along the major and intermediate 
principal axes of the large-scale tidal fields.  The clusters of galaxies are usually located at the junctions of multiple filaments~\cite{dur-etal16} that correspond 
to the zero-dimensional knots formed through the gravitational collapse along all of the three principal axes of the large-scale tidal fields~\cite{PS74,bon-etal91,mon95}. 
 
The sheets, the first large-scale structure ever formed through the collapse along the major principal axes of the tidal fields~\citep{sha-etal94}, mark the frontiers between 
the overdense and underdense regions, while the voids are the most underdense regions expanding rather than collapsing more rapidly than the rest of the 
universe~\cite{zel82,ick84,roo88}. 
Similar to the collapse process, the expansions of voids also occur anisotropically, leading them to develop non-spherical shapes stretched along the major principal directions 
of the large-scale tidal fields~\cite{sha-etal06,LP06,PL07}. In other words, the directions along which the overdense regions become most compressed coincide with those 
along which the neighboring underdense regions become most stretched~\cite{tru-etal06}. 

Numerous works have been devoted to understanding what role the cosmic web plays in establishing the galaxy properties
~\cite{hah-etal07,LL08,bam-etal09,alp-etal16,che-etal17,lee18,ala-etal19,kot-etal22,DP24,has-etal24,hoo-etal24,NP25}. 
Especially the alignments of galaxy shapes/spins with the filaments have recently become one of the hottest topics in the field of the large-scale structure, since these intrinsic 
alignments could contaminate the weak gravitational lensing signals ~\cite{align_review15a,align_review15c}.
It has been confirmed both numerically and observationally that the galaxy shapes are strongly aligned with the hosting filaments~\cite[][for a review]{align_review15b}, 
which are usually ascribed to the occurrence of preferential accretion of matter and merging of satellites along the filaments~\cite{zha-etal13,cam-etal15}. 
Regarding the galaxy spins (the directions of galaxy angular momenta), it has been found that they exhibit a mass-dependent transition of alignment tendency from being 
perpendicular to parallel to the filaments as the galaxy masses increase beyond a redshift-dependent threshold~\cite{ara-etal07,hah-etal07,BF12,lib-etal13,dub-etal14}.
In several literatures, it was claimed that this mass-dependent spin transition could also be explained by the consequence 
of filamentary merging~\cite{cod-etal12,dub-etal14,cod-etal18,gan-etal19,kra-etal20}.

It is not only the intrinsic alignments of galaxy shapes and spins that the filamentary merging along the cosmic web has been believed to account for. 
Several puzzling phenomena involving the galaxy properties like star formation quenching, gas-stripping, and secondary bias have been found to be at least partially resolved and 
understood if the effects of the cosmic web were properly taken into consideration~\citep{gao-etal05,GW07,FW10,zomg1,kuu-etal17,don-etal22,hoo-etal24}. 
In most of the previous works that studied the effects of the cosmic web on the galaxy properties, the main focus was usually put on the filament-galaxy connections, mainly  
because the filaments host the largest number of galaxies~\cite{ara-etal10,lib-etal18,gan-etal19}. Nevertheless, the galaxies located on less underdense sites like void surfaces 
should be optimal targets for sorting out the cosmic web effects, since the other effects from non-gravitational sources on the galaxy properties are likely to be less severe in those 
less crowded environments~\cite{tru-etal06,cue-etal08,var-etal12,kur-etal18,rez-etal19}.  

Furthermore, it has two additional advantages to investigate the cosmic web effects from the void-galaxy connections. 
First, for the case of the filaments, their elongated axes are usually determined as the minor principal axes of the local tidal fields, which in turn can be reconstructed from 
the spatial distributions of galaxies. The three dimensional reconstruction of the local tidal fields~\cite{erd-etal06,wan-etal12}, 
however, requires to make a simplified assumption about the galaxy bias factor, which quantifies how strongly the spatial correlations of galaxies differ from those of underlying matter. 
In dense environments like filaments, this galaxy bias factor has been known to deviate substantially from the simple linear form~\cite{nonlinear_bias1,nonlinear_bias2,lnl_bias}. 
Whereas, for the case of the voids, the major principal axes of the large-scale tidal fields can be readily determined in three dimensional space as the directions 
from the points on the void surfaces to the void centers~\cite{tru-etal06,cue-etal08}. 
This advantage turned out to be especially useful when the alignments between measurable three dimensional spin axes of spiral galaxies and principal directions of 
the local tidal fields are to be measured~\citep{LE07}. For instance, a significant signal of the mass-dependent transition of galaxy spin alignments was able to be detected directly 
from observations only when the alignments were measured on void surfaces~\cite{LM23}.

Second, for the case of the filaments, their effects on the galaxies sensitively depend on their geometrical features like their thicknesses~\cite{zomg1}, since they form 
through collapse only along two principal directions of the large-scale tidal fields. For instance, it was shown by ref.~\cite{zomg1} that if a filament hosts low-mass (high-mass) 
galaxies whose sizes are smaller (larger) than its thickness, it would play the role of promoting (repressing) infall of satellite and gas particles onto the galaxies by generating 
their radial (tangential) motions along the directions parallel (perpendicular) to the filaments. 
Whereas, for the case of the voids which form through expansion along all of three principal directions of the large-scale tidal fields, their effect was invariably only compression 
of adjacent matter without depending on their geometrical features. 

\begin{figure}[tbp]
\centering 
\includegraphics[width=0.85\textwidth=0 380 0 200]{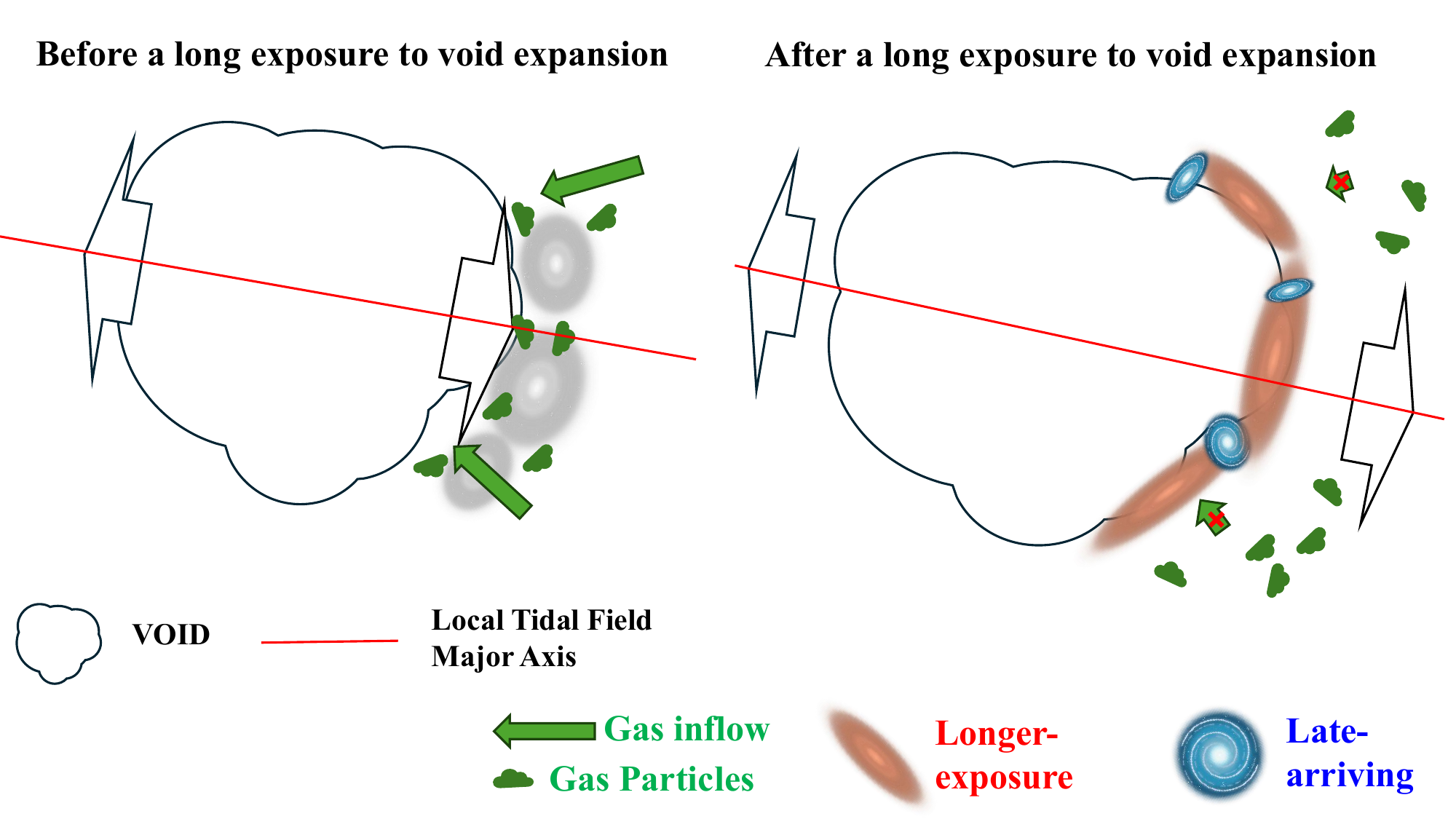}
\caption{\label{fig:cartoon} Schematic representation of two-fold effects of the void expansion on the void-surface galaxies. After a long exposure of 
a void-surface galaxy to the void expansion, it becomes stalled and quiescent, exhibiting strong perpendicular alignment of its shape axis with the 
direction to the void center. }
\end{figure}

In fact, the compression of adjacent matter caused by rapid expansion of a giant void is expected to have two-fold effects on the galaxies located on its surface. 
Figure~\ref{fig:cartoon} displays 
First, it would generate their tangential motions parallel to the directions from the void centers, which in turn would obstruct radial infall of gas particles onto the galaxies 
located on its surface. In consequence, provided that the adjacent galaxies are exposed to this compression effects for a long period of time, they would  experience gradual 
transformation from being accreting to stalled.
Second, it would elongate the shapes of galaxies on its surface in the directions perpendicular to the directions of maximum expansion, creating intrinsic perpendicular 
alignments of the galaxy shapes with the directions toward the void center. These two-fold effects of void expansions on the void-surface 
galaxies, illustrated in figure~\ref{fig:cartoon}, could be detected by exploring if and how the strengths 
of the void-surface galaxy perpendicular alignments differ between the stalled and accreting galaxies on void surfaces, which is the goal of this paper.  

Section~\ref{sec:analysis} is devoted to describing how the void-surface galaxy perpendicular alignments are numerically determined and how well the numerical 
results agree with an analytical one-parameter model. 
Section~\ref{sec:dep} is devoted to explaining how to create the stellar-mass controlled samples and describing how the alignment strengths from the controlled samples 
depend on the galaxy morphology, colors at three different redshifts. 
Section~\ref{sec:robust} is devoted to describing the robustness tests of the void-surface galaxy perpendicular alignments against the void-identification scheme, redshift-space 
and projection effects. 
Section~\ref{sec:con} is devoted to summarizing the main results and discussing a speculation of using the fraction of elliptical void-surface galaxies as a  complementary probe 
of dark energy (DE).
Throughout this paper, however, we assume a standard cosmology where the cosmological constant, $\Lambda$, accelerates the universe at the present epoch, and the most 
dominant matter content is the non-baryonic cold dark matter (CDM), and the spacetime is flat.

\section{Detection of the void-surface galaxy perpendicular alignments}\label{sec:analysis}

\subsection{Numerical data and analysis}\label{sec:procedure}
\begin{table}[tbp]
\centering
\begin{tabular}{cccc}
\hline
\hline
\rule{0pt}{4ex}\noindent
redshift & $N_{\rm v}$ & $\bar{R}_{\rm v}$ & $\bar{\delta}_{v}$ \\
 & & $[h^{-1}{\rm Mpc}]$ &  \\
\hline
0 &  $ 3123 $ & $8.88$ & $-0.89$  \\
0.5 &  $3187$ & $8.72$ & $-0.90$ \\
1 &  $ 3236$ & $8.61$ & $-0.90$  \\
\hline
\end{tabular}
\caption{\label{tab:void}
Total numbers, mean effective radii, mean density contrasts of the voids identified at three different redshifts via the Void-Finder algorithm.}
\end{table}
\begin{figure}[tbp]
\centering 
\includegraphics[width=0.85\textwidth=0 380 0 200]{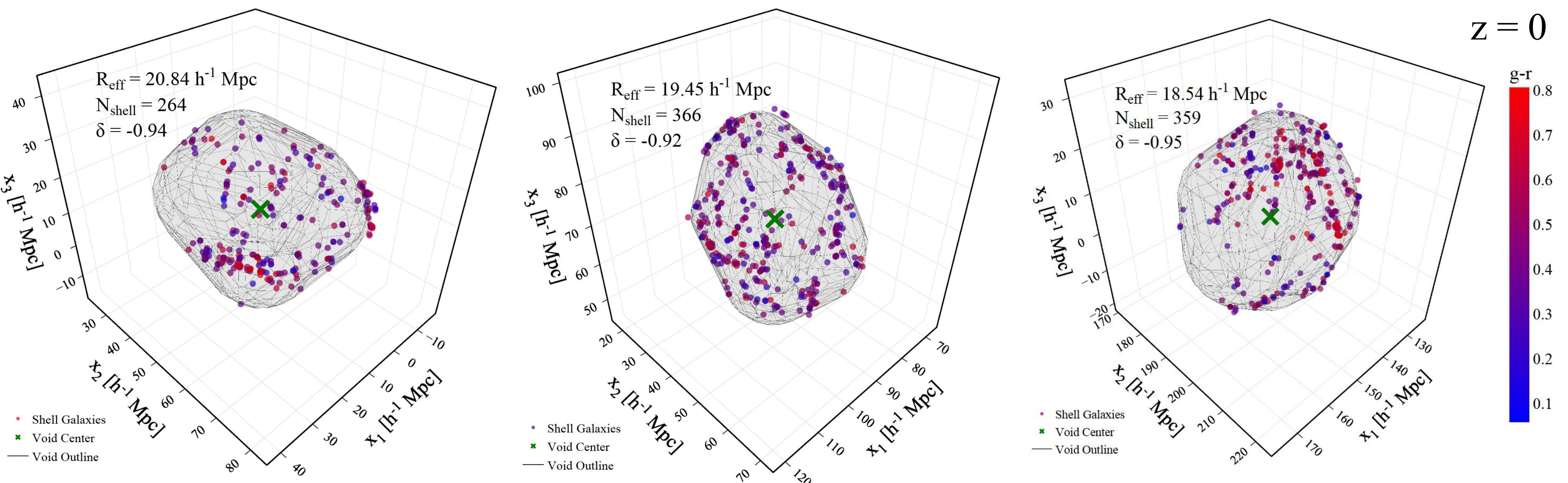}
\caption{\label{fig:void} Illustrations of the galaxies located on the surfaces of three voids, each being identified as an assembly of maximal spheres that fit large empty 
regions among the spatial distributions of the galaxies from the TNG 300-1 simulations~\cite{tng1,tng2,tng3,tng4,tng5,tng6} via the Void-Finder 
algorithm~\cite{HV02} at $z=0$.  The vertical bar displays the range and variation of the $g$-$r$ colors of the void-surface galaxies.}
\end{figure}
\begin{figure}[tbp]
\centering 
\includegraphics[width=0.85\textwidth=0 380 0 200]{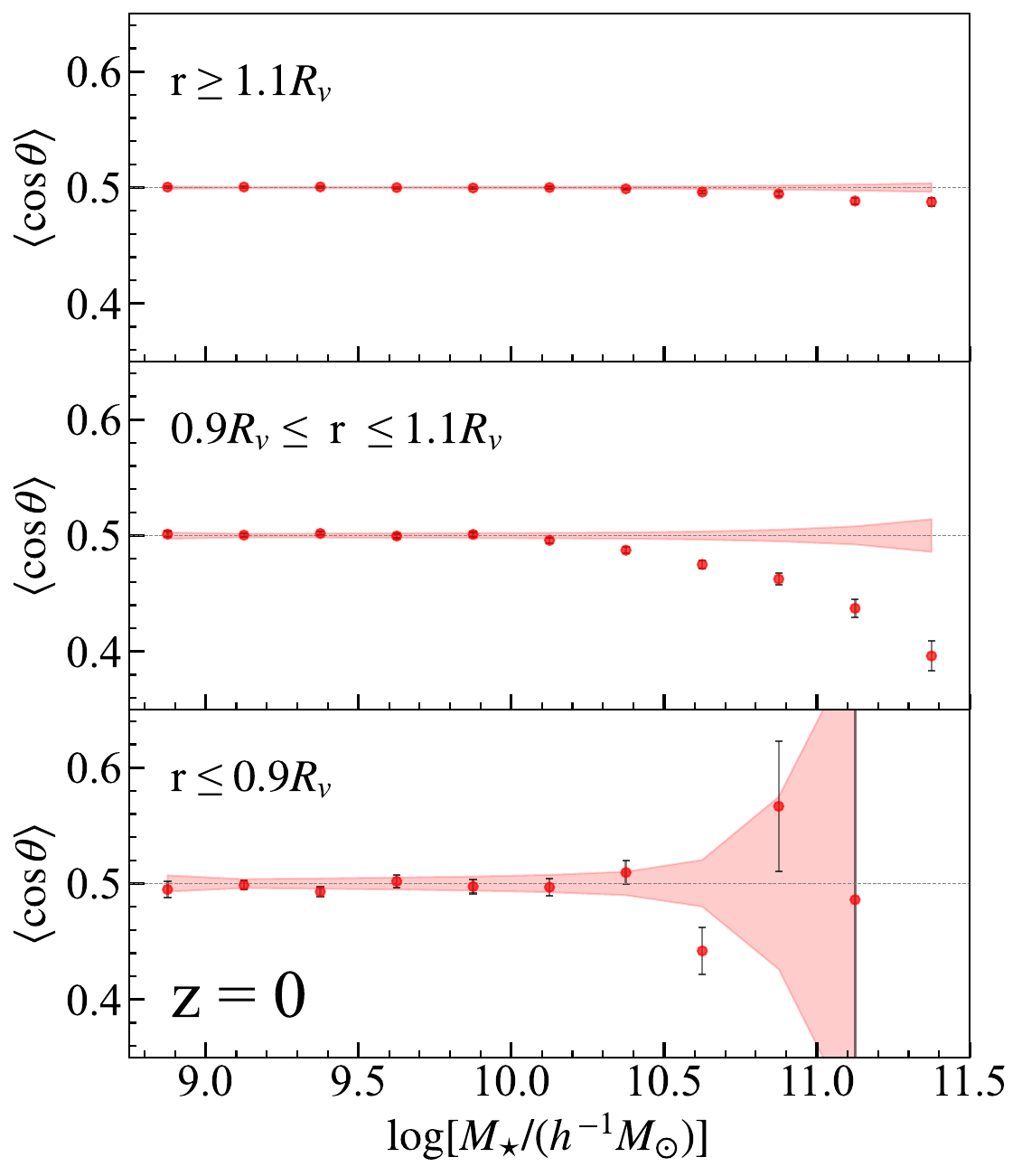}
\caption{\label{fig:mcost}  Mean values of the cosines of the angles between the shape axes of the TNG galaxies and the directions toward the 
void centers in three different ranges of separation distances, $r$, as a function of stellar masses }
\end{figure}
\begin{figure}[tbp]
\centering 
\includegraphics[width=0.85\textwidth=0 380 0 200]{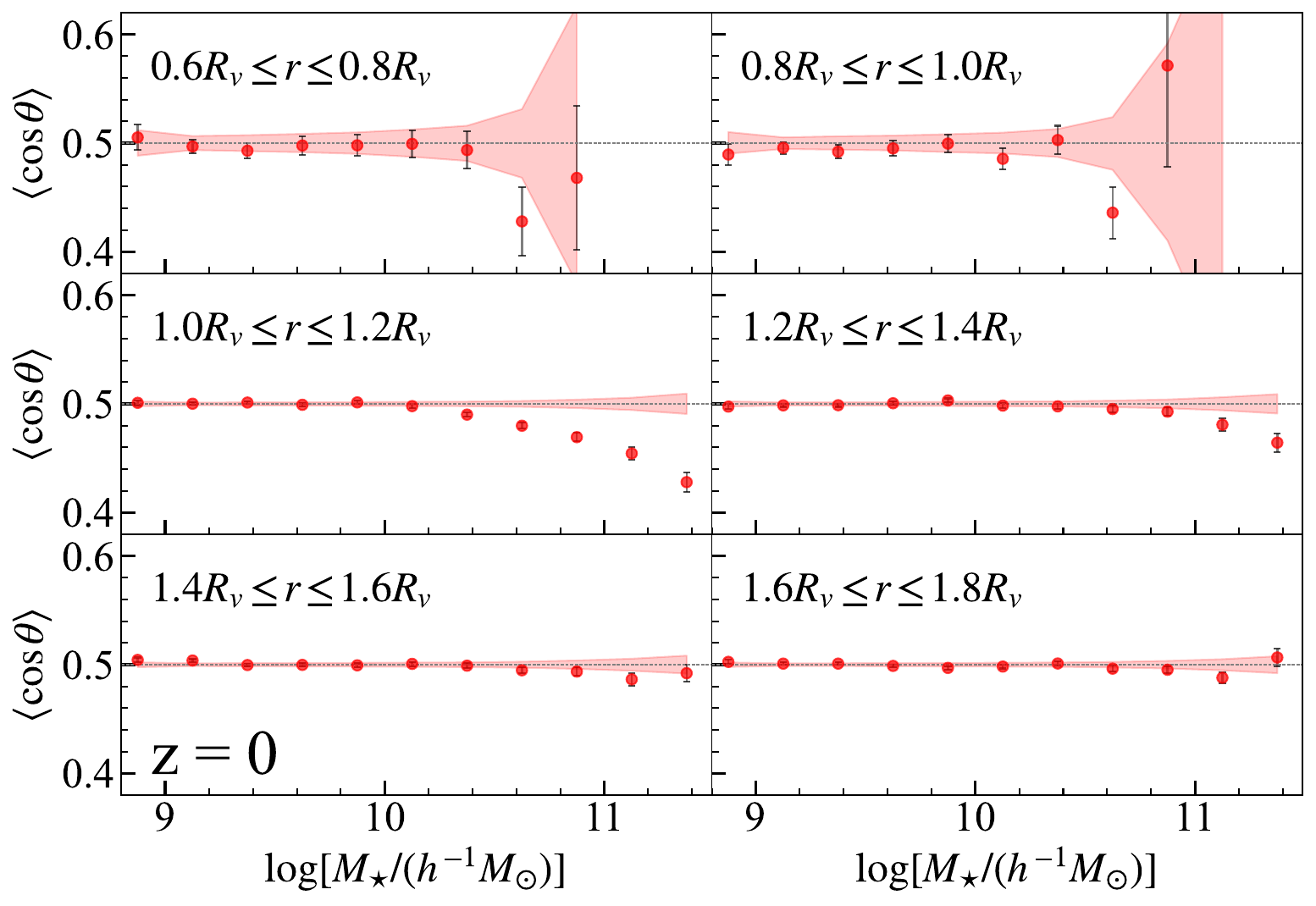}
\caption{\label{fig:multi} Same as figure~\ref{fig:mcost} but with the galaxies embedded in six different spherical shells of same thickness, 
$R_{\rm v}/5$ with void effective radii $R_{\rm v}$.}
\end{figure}
\begin{figure}[tbp]
\centering 
\includegraphics[width=0.85\textwidth=0 380 0 200]{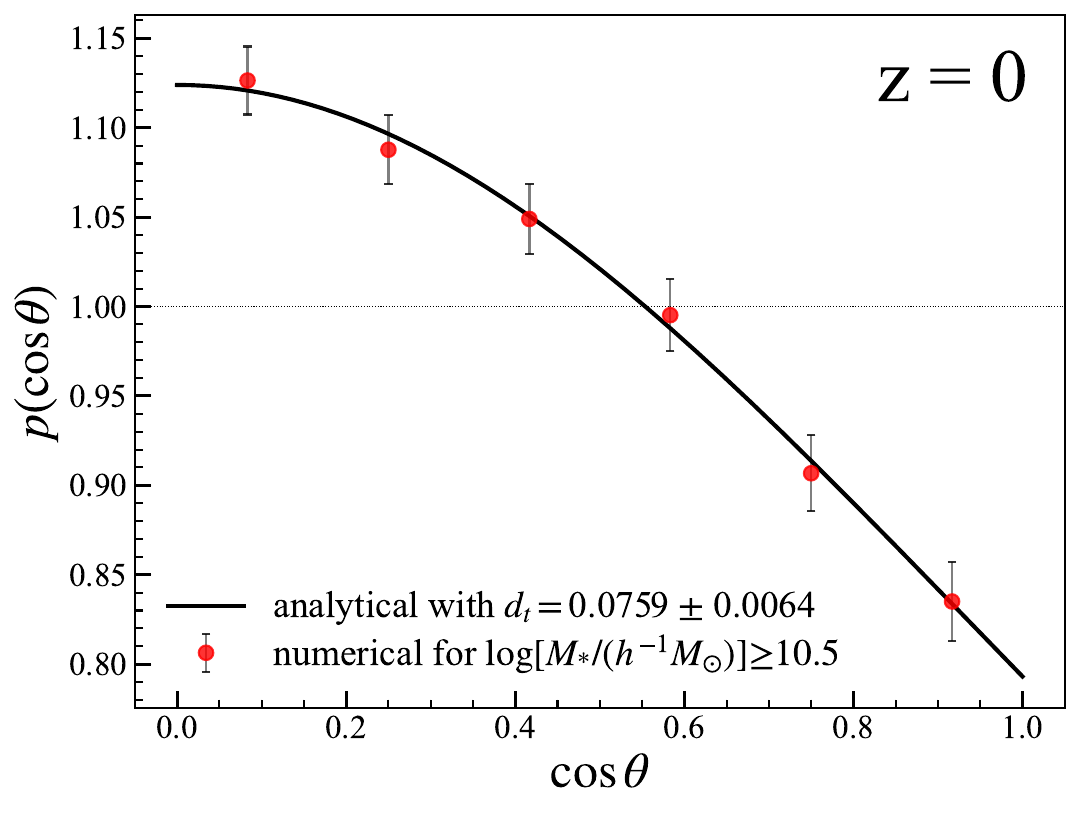}
\caption{\label{fig:pcost}  Probability density function of the cosines of the angles with Poisson errors between the shape axes of the giant void-surface galaxies and the directions 
toward the void centers from the TNG 300-1 simulations, and its comparison with the best-fit analytic model, eq.~(\ref{eqn:pcost}). 
The best-fit value of the correlation parameter $d_{t}$ in eq.~\ref{eqn:eiej} is determined via $\chi^{2}$-minimization method and its associated error corresponds to the 
interval, $\Delta d_{t}$, over which the integral of the $\chi^{2}$-distribution satisfies the condition of $\int_{\Delta d_{t}} p(\chi^{2})d\chi^{2} \approx 0.683$.}
\end{figure}
To investigate the void-surface galaxy perpendicular alignments, we utilize the publicly available datasets of the 300 Mpc volume run (TNG300-1) of the IllustrisTNG suite of cosmological  
hydrodynamical simulations~\citep{tng1,tng2,tng3,tng4,tng5, tng6}, the initial conditions of which were set at the Planck values~\cite{planck15}, assuming a flat 
$\Lambda$CDM cosmology. The TNG300-1 simulation keeps track of $2\times 2500^{3}$ particles starting from the initial redshift, $z=127$, to the present 
epoch, $z=0$, in a periodic cubic box of a side length $205\,h^{-1}{\rm Mpc}$ by implementing the AREPO moving-mesh code~\cite{arepo}.  
It has the same number fractions of DM and gas particles, whose mass resolutions are as high as $5.9$ and $1.1$ in unit of $10^{7}\,M_{\odot}$, respectively.   
The DM halos and their subhalos were resolved from the particle snapshots at various redshifts via the friends-of-friends~\cite{fof} and 
SUBFIND algorithms~\cite{subfind}, respectively, which are also publicly available at the TNG website\footnote{https://www.tng-project.org/data/}.  

Various information on each subhalo such as its total mass ($M$), stellar mass ($M_{\star}$), stellar formation epoch ($t_{f\star}$), colors and sSFR as well as on the positions and 
velocities of constituent stellar particles can be extracted from the catalog. Here, the stellar formation epoch of a subhalo is defined as the time when its stellar mass reaches a half of its 
present stellar mass. Selecting those well-resolved subhalos that have more than ten stellar cells, we determine the inertia tensor, 
$(I_{ij})$, of each subhalo's luminous part by using information on the positions of its stellar particles, as described in ref.~\cite{bet-etal07}.  
From here on, we refer to the selected subhalos as galaxies.
For each galaxy, a similarity transformation of $(I_{ij})$ is performed to find its three eigenvalues and corresponding eigenvectors. The {\it shape axis}, $\hat{\bf u}$, 
of each galaxy is then determined as the eigenvector of $(I_{ij})$ corresponding to the largest-eigenvalue.

We identify voids as large empty regions with effective sizes larger than a threshold, $s_{c}$, from the galaxy spatial distributions via the Void-Finder algorithm~\cite{HV02}. 
Those empty regions whose sizes smaller than threshold, $s_{c}$, are regarded as gaps among discretely distributed galaxies and thus excluded from the analysis. 
The value of $s_{c}$ is uniquely determined by the spatial distributions of the galaxies in a given catalog through a significance test. 
From the TNG 300-1 galaxy catalog at $z=0$, we find $s_{c}=5\,h^{-1}{\rm Mpc}$.  For a detailed description of the significance test through which the value of $s_{c}$ is 
determined, we refer the readers to our prior work~\cite{KL25} as well as to the original work of ref.~\cite{EP97} who first devised the significance test to discriminate true voids 
from mere gaps in the Poisson distributions of points. 

In the following is summarized the procedure that we follow to determine the void-surface galaxy perpendicular alignments:  
\begin{itemize}
\item
From each galaxy,  determine the separation distance, say $d_{3}$, to its third nearest neighbor. 
Taking the ensemble average, $\langle d_{3}\rangle$, and its  standard deviation,  $\sigma_{3}$, over all galaxies, 
determine the value of $l_{c}$ as $\langle d_{3}\rangle + 3\sigma_{3}/2$, which is found to be $l_{c}=2.82\,h^{-1}{\rm Mpc}$.
Dichotomize the galaxy into two categories according to the value of $l_{c}$: a wall galaxy ($d_{3}<l_{c}$) or a field galaxy ($d_{3}\ge l_{c}$).
\item
Partitioning the simulation box into multiple grids of equal side length $l_{c}$, find a block of empty grids which includes no wall galaxy and determine its 
best-fit sphere. Find a maximum empty sphere of each block, the farthest boundary of which just begins to reach three wall galaxies
\footnote{In the original work of ref.~\cite{HV02} which developed the Void-Finder algorithm, the terminology "wall" was used to refer not to the sheets but to all dense environments 
including the sheets, filaments and knots.}. 
\item
Search for the largest empty sphere in the simulation box and mark it as a maximal one.  Then, find the second largest empty sphere in the box and 
mark it as another maximal one, if its volume intersects with the largest one by less than $10\%$. 
Iterate this search with the other empty spheres of smaller volumes in a decreasing order, as far as their radii are larger than $s_c$.
\item 
Combine the non-maximal spheres into their nearest maximal if their volumes intersect with that of their nearest maximal by more than $50\%$. 
This combined region composed of the maximal and its intersecting non-maximal spheres is finally identified as a void. 
\item
With the help of the Monte-Carlo integration method~\cite{LP06}, determine the volume, $V_{\rm v}$, of each void. Then, determine 
its effective radius as $R_{\rm eff}\equiv \left[3\,V_{\rm v}/(4\pi)\right]^{1/3}$.
\item
For each void identified as an assembly of several empty spheres, look for the galaxies located on the surfaces of the constituent empty-spheres with radii 
$R_{\rm v}$. Find the galaxies that satisfy the condition of $0.9\le r/R_{\rm v}\le 1.1$ where $r$ is the separation distance from the center of its nearest neighbor 
empty sphere that belongs to a given void to the galaxy position. These galaxies will be regarded as being located on the surface of a given void, and called 
the void-surface galaxies from here on. Figure~\ref{fig:void} illustrates three examples of the voids (shaded volumes) identified in the TNG 300-1 simulations at $z=0$ 
and their surface galaxies (filled dots). The first row of table~\ref{tab:void} provides information on the total number, mean effective radius, and mean density contrast of 
the voids identified via the Void-Finder algorithm at $z=0$. 
\item
For each galaxy on the surface of a given void, determine a unit separation vector, $\hat{\bf r}$, pointing from its position toward the void center. 
This separation vector will be regarded as the major principal axis of the local tidal field at the position of a given galaxy on the void surface. 
Then, compute $\cos\theta\equiv \vert\hat{\bf r}\cdot\hat{\bf u}\vert$.
\end{itemize}

Splitting the ranges of $m_{\star}\equiv \log\left[M_{\star}/(h^{-1}M_{\odot})\right]$ into multiple short intervals, we take the ensemble average, $\langle\cos\theta\rangle$, 
over the void-surface galaxies belonging to each $m_{\star}$-interval to see how the strengths of void-surface galaxy perpendicular alignments vary with 
$m_{\star}$, the results of which are shown in the middle panel of figure~\ref{fig:mcost}. The errorbars represent one standard deviation in  $\langle\cos\theta\rangle$, 
while the shaded area correspond to $1$ $\sigma$ scatters of $\langle\cos\theta\rangle$ among $10,000$ resamples created by randomly shuffling the 
positions of the void-surface galaxies. A clear signal of the void-surface galaxy perpendicular alignments, $\langle\cos\theta\rangle<0.5$, is detected from the void-surface 
galaxies, becoming stronger in the higher-$m_{\star}$ section. 

To see whether or not the galaxies inside or outside the voids also exhibit this perpendicular alignment tendency, we refollow the above procedure but with the 
galaxies located at distances larger and smaller than the range of $0.9\le r/R_{\rm v}\le 1.1$, the results of which are shown in the top and bottom panels of figure~\ref{fig:mcost}, 
respectively. As can be seen, even in the high-$m_{\star}$ sections, the mean values, $\langle\cos\theta\rangle$, seem to be quite consistent with the constant $0.5$, 
the value expected from random orientations of galaxy shapes. 
The results shown in figure~\ref{fig:mcost} indicate that only the void-surface galaxies in the range of $0.9\le r/R_{\rm v}\le 1.1$ with $m_{\star}\ge 10.5$ exhibit significant signals 
of the void-surface galaxy perpendicular alignments. 

To confirm that the perpendicular alignment tendency is a unique phenomenon exhibited only by those galaxies located in the thin spherical shells surrounding the voids, 
we divide the volume of each void and its neighbor surrounding regions into $six$ spherical shells of same thickness of $R_{v}/5$. Locating the galaxies whose radial distances 
fall in each spherical shell, we compute their alignments with the directions toward the void center and then take the average over all voids, the results of which are shown 
in figure~\ref{fig:multi}. As can be seen, significant signals of the perpendicular alignments are found only from those spherical shells in the radial ranges of $1\le r/R_{\rm v}\le 1.4$.
Note also that a much stronger signal of the perpendicular alignments is found in the range of $1\le r/R_{\rm v}\le 1.2$ than in $1.2\le r/R_{\rm v}\le 1.4$. This result provides 
a strong support to our claim that the perpendicular alignments with the directions toward the void centers are indeed a unique properties of the void-surface galaxies.

The probability density function of $\cos\theta$ is determined as $p(\cos\theta)\equiv n_{g}/(N_{\rm tot}\,d\cos\theta)$ from the giant void-surface galaxies with  $m_{\star}\ge 10.5$. 
Here, $n_{g}$ denotes the number of the giant void-surface galaxies, whose values of $\vert\hat{\bf u}\cdot\hat{\bf r}\vert$ fall in a short bin of $[\cos\theta,\cos\theta+d\cos\theta]$ 
of equal length $d\cos\theta$, while $N_{\rm tot}$ denotes the total number of the giant void-surface galaxies with $m_{\star}\ge 10.5$. The uncertainty in the determination of 
$p(\cos\theta)$ is computed as $1\sigma$ Poisson error as $\sigma = 1/\sqrt{n_{g}-1}$ since $p(\cos\theta)$ becomes unity if there is no alignment tendency. 
Figure~\ref{fig:pcost} displays this numerically determined $p(\cos\theta)$ (red filled circles) with Poisson errors.  As can be seen, a clear trend of monotonic decrease of $p(\cos\theta)$ with 
$\cos\theta$ is found,  confirming that the shapes of the void-surface galaxies are indeed perpendicularly aligned with the directions to the void centers. This result turns out to be robust against the 
variation of the thickness of the shells around the void surfaces, $\Delta r/R_{\rm v}$. The $10\%$ increment (decrement) of $\Delta r/R_{\rm v}$ is found to yield only $2\%$ change in the value of $d_{t}$. 

\subsection{Comparison with the analytic model}\label{sec:model}
\begin{figure}[tbp]
\centering 
\includegraphics[width=0.85\textwidth=0 380 0 200]{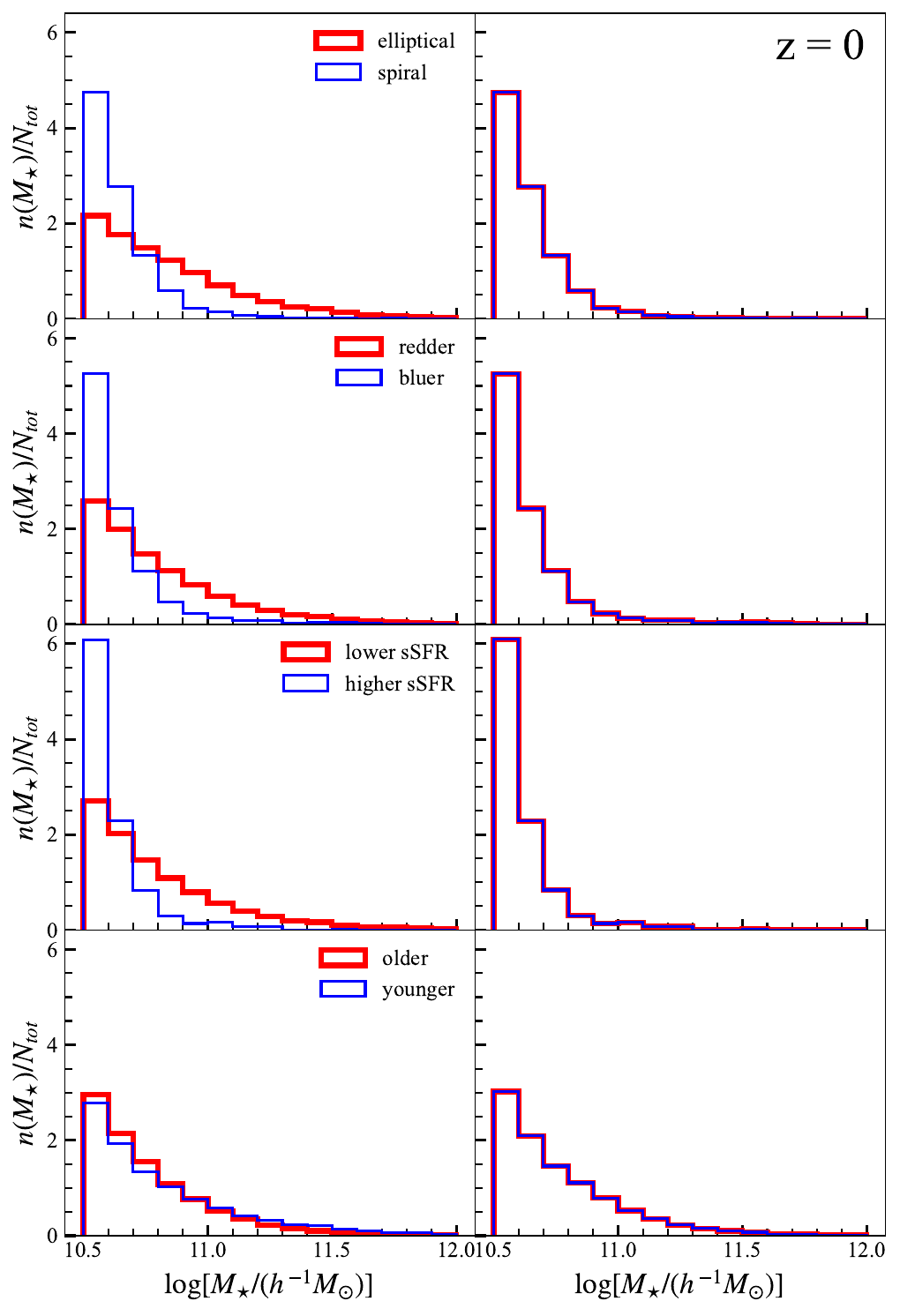}
\caption{\label{fig:ori_con} (Left panels): Number fractions of the galaxies from two samples, original (left panels) and $m_{\star}$-controlled (right panels), 
dichotomized by four different galaxy properties that separate the stalled galaxies from the accreting ones at $z=0$. Only those giant void-surface 
galaxies satisfying $m_{\star}\ge 10.5$ and $0.9\le r/R_{\rm v}\le 1.1$ are included for these plots.}
\end{figure}
\begin{figure}[tbp]
\centering 
\includegraphics[width=0.85\textwidth=0 380 0 200]{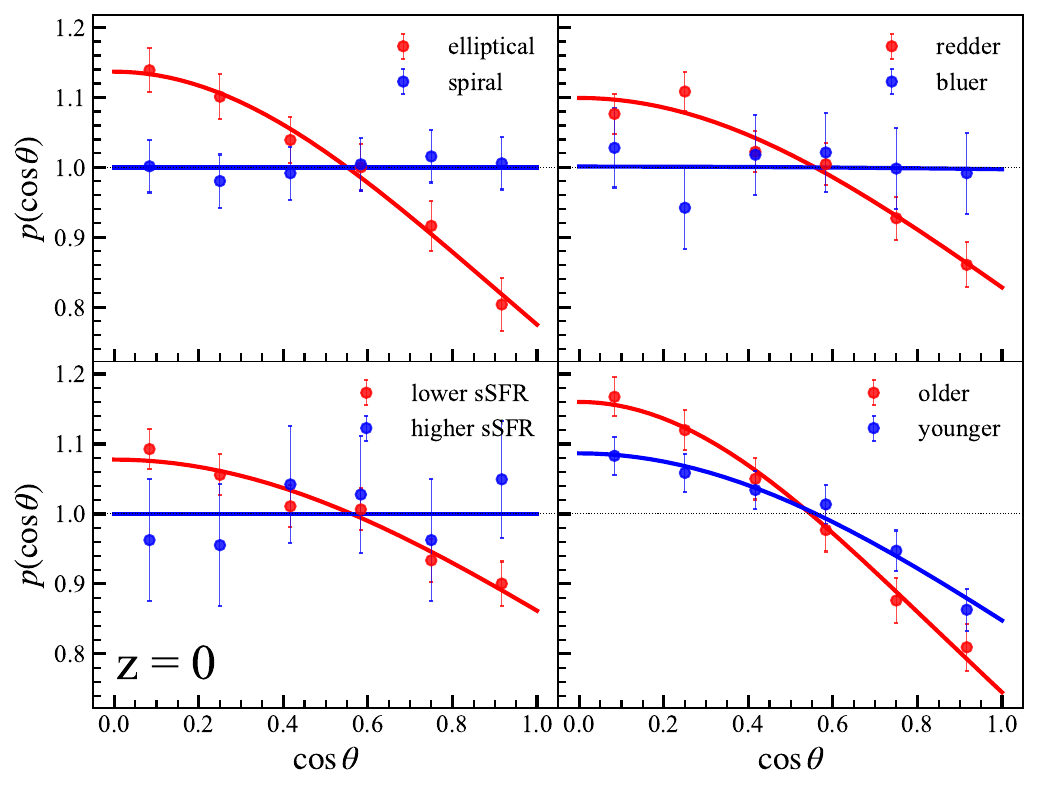}
\caption{\label{fig:pcost_dep} Dependences of $p(\cos\theta)$ with Poisson errors on the four galaxy properties. 
Only those giant void-surface galaxies satisfying $m_{\star}\ge 10.5$ and $0.9\le r/R_{\rm v}\le 1.1$ are included for these plots. }
\end{figure}
\begin{figure}[tbp]
\centering 
\includegraphics[width=0.85\textwidth=0 380 0 200]{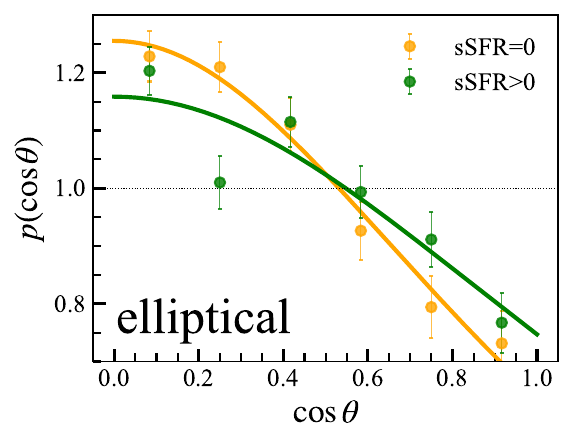}
\caption{\label{fig:mor_fix} Differences in $p(\cos\theta)$ between the star-forming  (${\rm sSFR}>0$) and quenched (${\rm sSFR}=0$) elliptical galaxies on void surfaces. 
The criterion of $\log{\rm sSFR}<-15$ adopted in the TNG 300-1 subhalo catalog is used to define the quenched galaxies.}
\end{figure}
\begin{figure}[tbp]
\centering 
\includegraphics[width=0.85\textwidth=0 380 0 200]{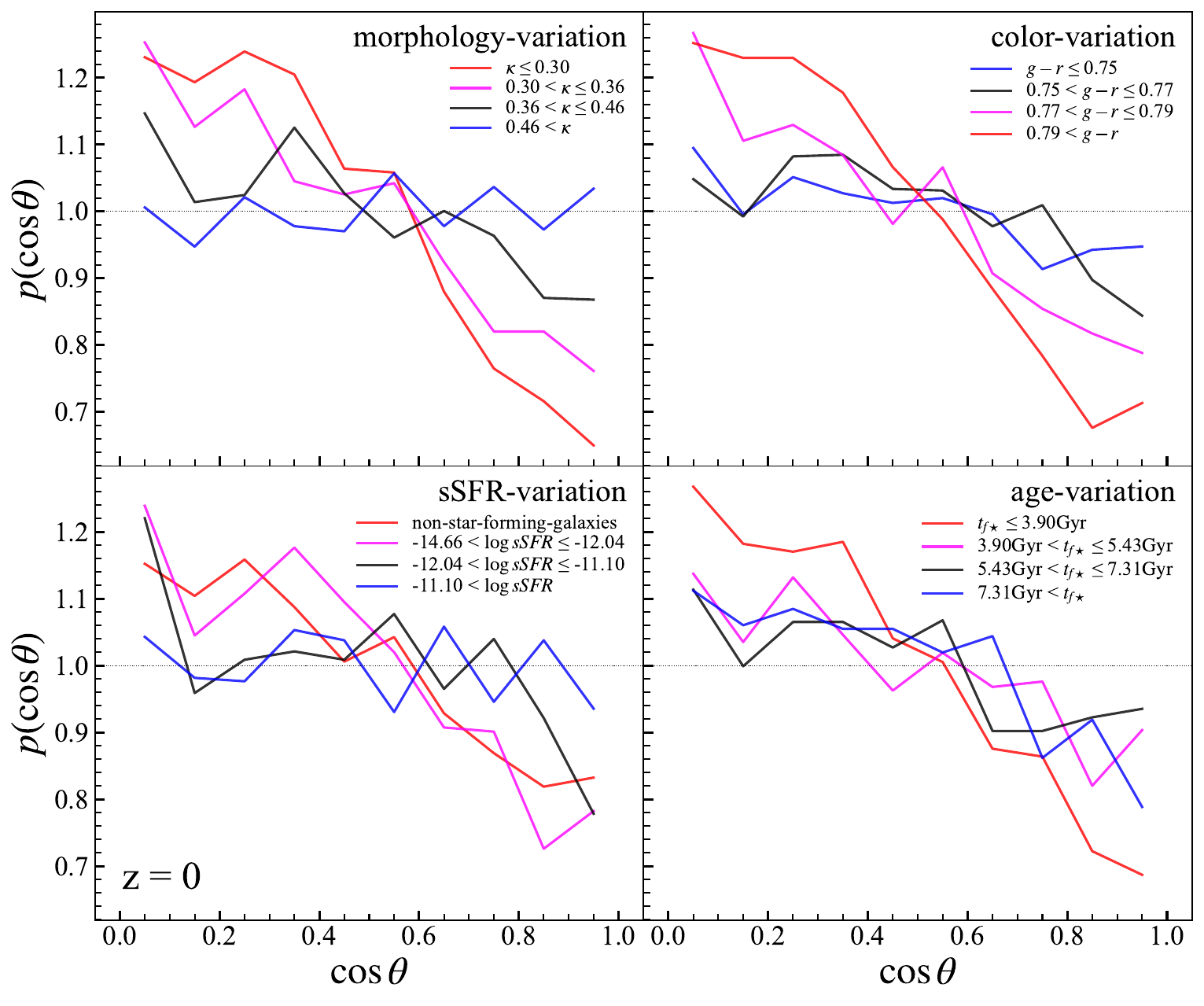}
\caption{\label{fig:pcost_var} Over-all trends of the variations of the void-surface galaxy perpendicular alignments with the four galaxy properties. Only those giant void-surface 
galaxies satisfying $m_{\star}\ge 10.5$ and $0.9\le r/R_{\rm v}\le 1.1$ are included for these plots.}
\end{figure}
\begin{table}[tbp]
\centering
\begin{tabular}{lccccccc}
\hline
\hline
\rule{0pt}{4ex}\noindent
criterion & $N_{\rm ori}$ & $N_{\rm cont}$ & $10^{2} d_{t}$ & $\langle \cos \theta \rangle$ & $p_{\rm ks}$ \\
\hline
\rule{0pt}{4ex}\noindent
$\kappa < 0.45$ & $30032$ & $13613$ & $8.45 \pm 1.04$ & $0.47 \pm 0.00$ &$<0.01$\\
$\kappa \ge 0.45$ & $11273$ & $11272$ & $0.00 \pm 0.34$ & $0.50 \pm 0.00$ &$1.00$\\
\hline
\rule{0pt}{4ex}\noindent
$g-r \ge 0.68$ & $37045$ & $18218$ & $6.12 \pm 0.90$ & $0.48 \pm 0.00$ &$<0.01$\\
$g-r < 0.68$ & $4260$ & $4256$ & $0.00 \pm 0.83$ & $0.50 \pm 0.01$ &$1.00$\\
\hline
\rule{0pt}{4ex}\noindent
$\log(\rm sSFR) < -10.4$ & $39422$ & $17518$ & $4.75 \pm 0.91$ & $0.48 \pm 0.00$ &$<0.01$\\
$\log(\rm sSFR) \ge -10.4$ & $1883$ & $1879$ & $0.00 \pm 0.73$ & $0.51 \pm 0.01$ &$0.98$\\
\hline
\rule{0pt}{4ex}\noindent
$t_{f\star} < 5.43\,{\rm Gyr}$ & $20021$ & $19478$ & $9.58 \pm 0.95$ & $0.46 \pm 0.00$ &$<0.01$\\
$t_{f\star} \ge 5.43\,{\rm Gyr}$ & $21261$ & $19478$ & $5.45 \pm 0.85$ & $0.48 \pm 0.00$ &$<0.01$\\
\hline
\end{tabular}
\caption{\label{tab:para}
Criterion used for the segregation, numbers of the void-surface galaxies in the original and  $m_{\star}$-controlled samples, 
best-fit parameter of the analytic model, mean void-surface galaxy alignment strengths and $p$-value of the KS test at $z=0$.}
\end{table}
To physically explain the numerical results presented in section~\ref{sec:procedure}, we adopt the analytic model developed by ref.~\cite{lee19} for the perpendicular alignments 
between the shape axes of DM halos and the major principal axes of the local tidal fields.
The key assumption of this model is that the principal axes of the halo inertia tensors tend to be anti-aligned with those of the local tidal fields, which is equivalent to assuming 
that the conditional covariance among the eigenvector corresponding to the largest-eigenvalue, $\hat{\bf u}=(\hat{u}_{1},\hat{u}_{2},\hat{u}_{3})$,  scales linearly with the 
unit traceless tidal tensor, $(\hat{T}_{ij})$ as
\begin{equation}
\label{eqn:eiej}
\langle\hat{\bf u}_{i}\hat{\bf u}_{j}\vert\hat{\bf T}\rangle = \frac{1+d_{t}}{3}\delta_{ij} - d_{t}\hat{T}_{ij}\, ,\,\,\, {\rm for}\,\,\, i,j\in\{1,2,3\}\, ,
\end{equation}
where $d_{t}$ is a free parameter to quantify the alignment strength, which has to be empirically determined~\cite{lee19}. 

Assuming further that the conditional probability distribution, $p(\hat{u}_{1},\hat{u}_{2},\hat{u}_{3}\vert\hat{\bf T})$, follows a multi-variate 
Gaussian distribution and plugging eq.(\ref{eqn:eiej}) into it, Lee~\cite{lee19} derived $p(\cos\theta)$ as\footnote{In the original work of ref.~\cite{lee19}, there is a typo 
in the analytic expression of $p(\cos\theta)$. It has been corrected here.}
\begin{eqnarray}
\label{eqn:pcost}
p(\cos\theta)
&=& \frac{1}{2\pi}\left[\left(1+d_{t}\right)\left(1+d_{t}-3d_{t}/\sqrt{2}\right)\left(1+d_{t}+3d_{t}/\sqrt{2}\right)\right]^{-\frac{1}{2}}\times \\
&& 
\int_{0}^{2\pi}d\phi\left[\frac{\cos^{2}\theta}{1+d_{t}-3d_{t}/\sqrt{2}} + \frac{(1-\cos^{2}\theta)\cos^{2}\varphi}{1+d_{t}} + \frac{(1-\cos^{2}\theta)\sin^{2}\varphi}{1+d_{t}+3d_{t}/\sqrt{2}}\right]^{-\frac{3}{2}}\, ,
\end{eqnarray}
where $\varphi$ represents the angle in the two-dimensional space spanned by the major and intermediate principal axes of $(\hat{T}_{ij})$ 
between the projected shape axis of a galaxy and the minor principal axis of $(\hat{T}_{ij})$. 
With the help of the $\chi^{2}$-minimization method, we fit equation~(\ref{eqn:pcost}) to the numerically obtained $p(\cos\theta)$ from those giant void-surface galaxies by adjusting 
$d_{t}$, the result of which is shown as solid line in figure~\ref{fig:pcost}. As can be seen, the analytic model with best-fit value of $d_{t}$ indeed describes the numerical 
distribution quite well. 
This good agreement confirms that equation~(\ref{eqn:pcost}) can validly describe not only the shape alignments of DM halos as shown in ref.~\citep{lee19} but also those 
of the stellar constituents with the principal axes of the local tidal fields. It also supports the notion that the major principal axes of the local tidal fields are indeed normal to the void 
surfaces and parallel to the directions toward the void centers. 

\section{Dependence of alignment strengths on the galaxy properties}\label{sec:dep}

\subsection{Variation with morphology and color}\label{sec:mor_col}

Now that the giant void-surface galaxies are found to exhibit strong perpendicular alignments of their shapes with the directions toward the void centers, we would 
like to explore if there is any significant difference in the alignment strengths between the stalled and accreting galaxies. The stalled (accreting) galaxies are often 
characterized by their distinct properties such as elliptical (spiral) morphology, red (blue) colors, and lower (higher) sSFR and older (younger) stellar ages~\cite{zomg1}. 
We first investigate the morphology-dependence by dichotomizing the galaxies into ellipticals or spirals according to the fraction of rotational energy. 
Using information on the comoving positions and velocities of stellar particles belonging to each galaxy, we compute the rotational kinetic energy, $K_{\rm rot}$, 
as well as the total kinetic energy, $K_{\rm tot}$. 
Taking the ratio, $\kappa \equiv K_{\rm rot}/K_{\rm tot}$~\cite{rod-etal17,lee-etal21}, and using the morphology segregation threshold of $\kappa_{c}=0.45$ given in~\cite{PC05}, 
we separately determine $p(\cos\theta)$ of the ellipticals ($\kappa<\kappa_{c}$) and of the spirals ($\kappa>\kappa_{c}$).

To eliminate any effects caused by different $m_{\star}$-distributions between the ellipticals and spirals,  we control the two samples to have identical 
$m_{\star}$-distributions, and redetermine $p(\cos\theta)$ from the $m_{\star}$-controlled samples of ellipticals and spirals. Here, $m_{\star}$-distributions are computed 
as $n(m_{\star})/N_{\rm tot}$ where $n(m_{\star})$ is the number of the void-surface galaxies with stellar masses in a differential interval 
of $[m_{\star},m_{\star}+dm_{\star}]$. 
The top panels of figure~\ref{fig:ori_con} plot the $m_{\star}$-distributions of the ellipticals (red histograms) and spirals (blue histograms),
confirming that the two controlled samples indeed have no difference in the $m_{\star}$-distributions (top-right panel) from each other, even though considerable differences 
in the $m_{\star}$-distributions are existent in the original samples (top-left panel). The numbers of the dichotomized void-surfaces galaxies in the original and 
$m_{\star}$-controlled samples are listed in the second and third columns of table~\ref{tab:para}. 

The top-panel of figure~\ref{fig:pcost_dep} shows $p(\cos\theta)$ from the $m_{\star}$-controlled samples of the ellipticals and spirals (filled red and blue circles, respectively) and 
the analytic model with the best-fit parameters (solid lines).  As can be seen, the spiral galaxies yield almost completely uniform $p(\cos\theta)$, while the elliptical galaxies 
exhibit quite strong signals of perpendicular shape alignments with the directions toward the void centers, which indicates that the void-surface galaxy perpendicular alignments are 
indeed dependent on the galaxy morphology. Note also that the elliptical galaxies exhibit a good match between the numerically obtained $p(\cos\theta)$ and the analytic model 
with the best-fit parameter (see the fourth column of table~\ref{tab:para}). 
We also perform the Kolmogorov-Smirnov (KS) test of the null hypothesis that there is no void-surface galaxy perpendicular alignments to compute the $p$-value 
(see the seventh column of table~\ref{tab:para}), which turns out to be rejected at the confidence levels higher than $99\%$ for the case of the elliptical galaxies. 
It is worth emphasizing here that this morphology dependence of the void-surface galaxy perpendicular alignments cannot be ascribed to any stellar-mass difference between 
the ellipticals and spirals since the signals are found from the $m_{\star}$-controlled samples.

We also investigate the color dependence of void-surface galaxy perpendicular alignments, performing a dichotomy of the void-surface galaxies according to their $g-r$ colors~\cite{tngcolor}. 
Since here is no established threshold of $g-r$ color unlike the case of morphology segregation, we first inspect if and how the strengths of the void-surface galaxy perpendicular 
alignments change with the $g$-$r$ colors and find that the void-surface galaxies with higher $g$-$r$ colors exhibit stronger signals. Then, we dichotomize the void-surface galaxies 
into redder ($g-r\ge 0.68$) or bluer ($g-r< 0.68$), where the threshold value of $0.68$ is found as the one below which no signal of the void-surface galaxy perpendicular alignments is 
found. Creating two $m_{\star}$-controlled samples of redder and bluer galaxies in the same manner as done for the morphology-segregation, we separately determine 
$p(\cos\theta)$. Figure~\ref{fig:ori_con} shows in its second panels from the top how well the samples of the redder and bluer void-surface galaxies are controlled to yield 
the identical $m_{\star}$-distributions, compared with the original ones. 
The top-right panel of figure~\ref{fig:pcost_dep} shows the differences in $p(\cos\theta)$ between the $m_{\star}$-controlled samples of the redder and bluer galaxies. 
Note that a significant signal of the void-surface galaxy perpendicular alignment is found only from the redder galaxies and that the analytic model, eq.~(\ref{eqn:pcost}), 
with the best-fit parameter (solid lines) agrees with the numerically derived $p(\cos\theta)$ for the redder case. 
In contrast, the distribution $p(\cos\theta)$ from the bluer galaxies seems to be statistically consistent with being uniform $p(\cos\theta)=0.5$, implying the 
existence of color-dependence of the void-galaxy perpendicular  alignments. Table~\ref{tab:para} lists a summary of these results for the color-dependence. 

\subsection{Variation with sSFR and stellar age}\label{sec:ssfr_age}

Since the stalled and accreting galaxies also differ from each other in their sSFR and stellar ages, we proceed to explore if the void-surface galaxy perpendicular alignments also depend 
on the sSFR and stellar formation epochs.  To dichotomize the giant void-surface galaxies into the relatively more quiescent or star-forming ones, we first find the threshold sSFR by taking 
the following steps. Using a median sSFR value as the first trial threshold, we split the galaxies into the higher and lower sSFR subsamples. 
Investigating how $p(\cos\theta)$ differs between the two subsamples, we iteratively change the threshold sSFR from the trial value until $p(\cos\theta)$ 
from the higher-sSFR subsample becomes uniform (i.e., constant).  Among the threshold sSFR values that meets the condition of $p(\cos\theta)={\rm constant}$ 
for the case of the higher-sSFR subsample, we choose the maximum as the final threshold. 

Another dichotomy is performed according to the galaxy stellar formation epochs, $t_{f\star}$, into the older or younger. For this segregation, we use the median values of 
$t_{f\star}$ obtained from the original samples as a threshold, since no value of $t_{f\star}$ is found to satisfy the condition that only one of the two samples segregated 
by $t_{f\star}$ yields a signal of the void-surface galaxy perpendicular alignments.   Then, we repeat the whole process described in section~\ref{sec:mor_col} with these segregated 
galaxy samples to separately determine, $p(\cos\theta)$, $d_{t}$, $\langle\cos\theta\rangle$ and the $p$-value of the KS test (see table~\ref{tab:para}). 

The bottom panels of figure~\ref{fig:pcost_dep} plots the same as the top-panels but for the cases that the galaxies are segregated by their values of sSFR (left panels) 
and by their stellar ages $t_{f\star}$ (right panels). As can be seen, the more quiescent and older galaxies yield significant signals of void-surface galaxy perpendicular alignments 
while the less quiescent and younger galaxies yield insignificant and weaker tendencies, respectively.  For all of the cases, the analytic model, eq.~(\ref{eqn:pcost}), 
succeeds in matching the numerical results, confirming its efficiency. 
These results are quite consistent with those presented in section~\ref{sec:mor_col}, confirming that the stalled galaxies characterized by elliptical shapes, redder colors, 
lower-sSFR and older stellar ages exhibit significantly stronger perpendicular alignments of their shape axes with the directions toward the void centers. 

Notwithstanding, we note a subtle difference among the four cases in each of which a different criterion is used to dichotomize the void-surface galaxy population. 
The two segregated samples yield the largest (smallest) difference in the strengths of void-surface galaxy perpendicular alignments if the segregation criterion is based on the 
galaxy morphology (stellar ages). In other words, the strengths of the void-surface galaxy perpendicular alignments exhibit the strongest (weakest) dependence on the morphology  
(stellar age) of the void-surface galaxies. Given this difference, we speculate that the elliptical, redder and more quiescent void-surface galaxies exhibit stronger perpendicular alignments 
not because they formed earlier but because they may be exposed for longer time to the expansions of voids which could contribute to quenching them.

At this stage, it is important to confirm that the observed stronger signals of elliptical galaxies are caused truly by an independent morphology-dependence of alignment strengths rather than by 
different quenching histories and galaxy ages among the void-surface galaxies. For this, we classify the elliptical void-surface galaxies into the star-forming and quenched ones, employing the 
criterion of $\log{\rm sSFR}=-15$ in accordance with the TNG 300-1 simulations, and investigate if and how the alignment strengths differ between them, the results of which 
are shown in figure~\ref{fig:mor_fix}. 
As can be seen, the star-forming ($\log{\rm sSFR}>-15$) and quenched ($\log{\rm sSFR}<-15$) elliptical galaxies on void-surfaces show no significant difference in the alignment strengths, albeit 
slightly stronger signals are found from the latter. Regarding the disk void-surface galaxies,  they are found to yield no significant signal of perpendicular alignments, regardless of their sSFR 
values. Furthermore, no quenched disk galaxies with $\log{\rm sSFR}<-15$ are found on void surfaces. These results support our claim that the void-surface galaxy perpendicular alignments 
indeed possess a strong {\it independent} dependence on the galaxy morphological type. Nevertheless, it is worth stressing here that  the absence of quenched disk galaxies on void surfaces and the 
lack of perpendicular alignments for disk galaxies should not be interpreted as evidence that void-driven processes transform disk galaxies into ellipticals.

To examine whether or not the void-surface galaxy perpendicular alignments exhibit systematic and monotonic variations with the aforementioned four properties, 
we split the range of each property into multiple bins and repeat separately the whole process with the galaxies belonging to each of the four bins, the results of which 
are shown in figure~\ref{fig:pcost_var}.  A consistent and systematic variation of $p(\cos\theta)$ with each of the four properties is witnessed.  The probability density function 
$p(\cos\theta)$ exhibits a trend of decreasing more rapidly with $\cos\theta$, as the fraction of rotational energy and sSFR gradually decrease and as the $g$-$r$ color 
and stellar ages $t_{f\star}$ increase. Note again that $p(\cos\theta)$ exhibits the mildest variation with $t_{f\star}$. 

It is worth comparing the results on the fractions of blue star-forming galaxies on void surfaces between the current and previous works. 
The previous work of ref.~\cite{cer-etal08} found a tendency that the "void-wall" galaxies tend to have bluer colors and higher sSFR than the filament counterparts, 
which is apparently inconsistent with the current result (see table~\ref{tab:para}) that the void-surface galaxies tend to be redder and quiescent. 
We suspect that this inconsistency is likely to be caused by the difference in the void-identification scheme.  In ref.~\cite{cer-etal08}, a void region was identified as a maximum sphere 
with its center at the local density minima, while in the current work it is identified as an assembly of overlapped empty spheres via the Void-Finder. 
From a perspective of the Void-Finder, the spherical surface that ref.~\cite{cer-etal08} called a "void-wall" is not the surface surrounding a void but some part of the 
sheet-like structure (i.e., mini-sheet) embedded in the void. As shown in ref.~\cite{mini_sheet}, even though the voids are characterized by the lowest densities, they contain plenty of mini-sheets 
and mini-filaments. Besides, it was observationally found by ref.~\cite{void_slow} that the galaxies inside the voids tend to have bluer colors and lower sSFR since the stars in void galaxies 
evolve more slowly.  This observational result seems to be fully consistent with the finding of ref.~\cite{cer-etal08},  only provided that the void-wall galaxies in ref.~\cite{cer-etal08} actually 
correspond to the wall galaxies embedded in voids.

\subsection{Alignment signals at higher redshifts}\label{sec:evol}
\begin{table}[tbp]
\centering
\begin{tabular}{cccccccc}
\hline
\hline
\rule{0pt}{4ex}\noindent
${\rm criterion}$ & $N_{\rm ori}$ & $N_{\rm cont}$ & $10^{2} d_{t}$ & $\langle \cos \theta \rangle$ & $p_{\rm 3D}$ \\
\hline
\rule{0pt}{4ex}\noindent
$\kappa < 0.45$ & $23821$ & $12986$ & $8.10 \pm 1.16$ & $0.47 \pm 0.00$ &$<0.01$\\
$\kappa \ge 0.45$ & $14776$ & $12986$ & $1.69 \pm 1.06$ & $0.49 \pm 0.00$ &$0.54$\\
\hline
\rule{0pt}{4ex}\noindent
$g-r \ge 0.63$ & $25287$ & $13875$ & $8.39 \pm 1.15$ & $0.47 \pm 0.00$ &$<0.01$\\
$g-r < 0.63$ & $13310$ & $13306$ & $0.18 \pm 0.60$ & $0.50 \pm 0.00$ &$0.98$\\
\hline
\rule{0pt}{4ex}\noindent
$\log(\rm sSFR) < -10.2$ & $31248$ & $15363$ & $6.41 \pm 1.05$ & $0.48 \pm 0.00$ &$<0.01$\\
$\log(\rm sSFR) \ge -10.2$ & $7349$ & $7349$ & $0.00 \pm 0.26$ & $0.51 \pm 0.01$ &$0.62$\\
\hline
\rule{0pt}{4ex}\noindent
$t_{f\star} < 4.29\,{\rm Gyr}$ & $18578$ & $18198$ & $10.67 \pm 1.05$ & $0.46 \pm 0.00$ &$<0.01$\\
$t_{f\star} \ge 4.29\, {\rm Gyr}$ & $19982$ & $18200$ & $3.45 \pm 0.93$ & $0.49 \pm 0.00$ &$0.07$\\
\hline
\hline
\end{tabular}
\caption{\label{tab:para_z0.5}
Same as table~\ref{tab:para} but at $z=0.5$.}
\end{table}
\begin{figure}[tbp]
\centering 
\includegraphics[width=0.85\textwidth=0 380 0 200]{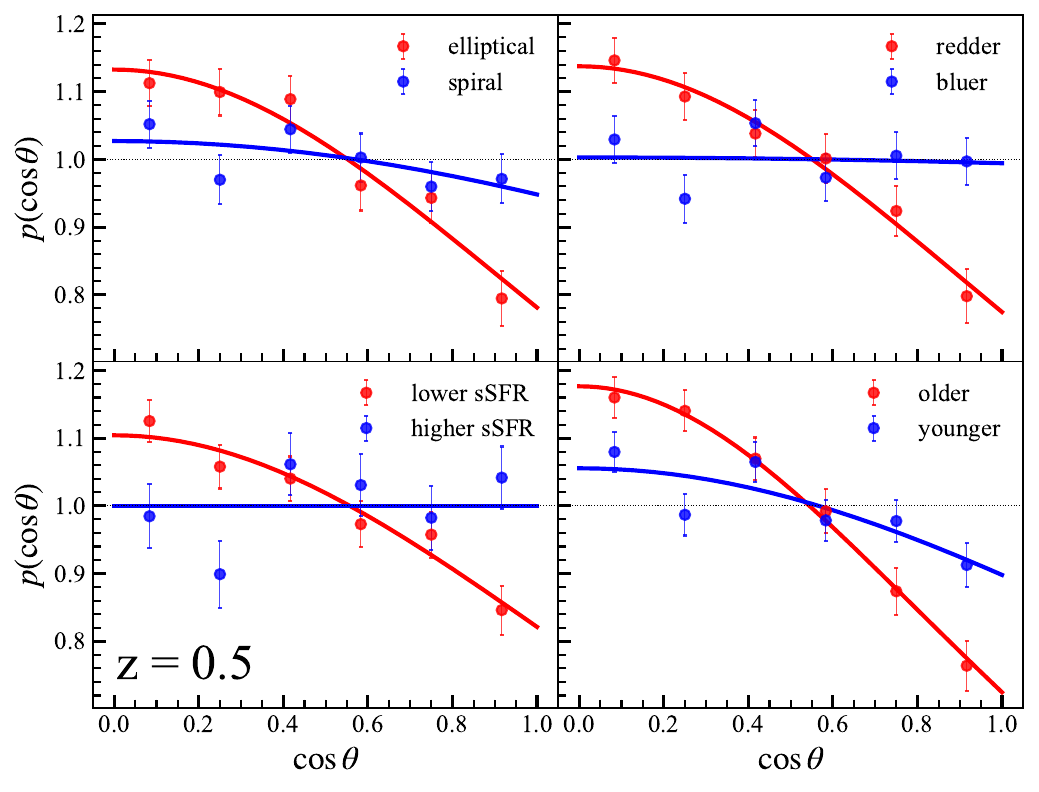}
\caption{\label{fig:post_dep_z0.5} Same as figure~\ref{fig:pcost_dep} but at $z=0.5$.}
\end{figure}
\begin{table}[tbp]
\centering
\begin{tabular}{cccccccc}
\hline
\hline
\rule{0pt}{4ex}\noindent
${\rm criterion}$ & $N_{\rm ori}$ & $N_{\rm cont}$ & $10^{2} d_{t}$ & $\langle \cos \theta \rangle$ & $p_{\rm KS}$ \\
\hline
\rule{0pt}{4ex}\noindent
$\kappa < 0.45$ & $16311$ & $10764$ & $9.75 \pm 1.35$ & $0.46 \pm 0.00$ &$<0.01$\\
$\kappa \ge 0.45$ & $14292$ & $10764$ & $0.38 \pm 0.81$ & $0.50 \pm 0.01$ &$0.70$\\
\hline
\rule{0pt}{4ex}\noindent
$g-r \ge 0.37$ & $24839$ & $12619$ & $4.20 \pm 1.19$ & $0.48 \pm 0.00$ &$0.05$\\
$g-r < 0.37$ & $5764$ & $5762$ & $0.45 \pm 1.04$ & $0.50 \pm 0.01$ &$0.93$\\
\hline
\rule{0pt}{4ex}\noindent
$\log(\rm sSFR) < -10.0$ & $21657$ & $10988$ & $5.42 \pm 1.30$ & $0.48 \pm 0.00$ &$0.03$\\
$\log(\rm sSFR) \ge -10.0$ & $8946$ & $8946$ & $0.57 \pm 0.94$ & $0.50 \pm 0.01$ &$0.76$\\
\hline
\rule{0pt}{4ex}\noindent
$t_{f\star} < 3.45\, {\rm Gyr}$ & $14696$ & $14143$ & $5.78 \pm 1.21$ & $0.48 \pm 0.00$ &$<0.01$\\
$t_{f\star} \ge 3.45\, {\rm Gyr}$ & $15880$ & $14143$ & $6.14 \pm 1.16$ & $0.48 \pm 0.00$ &$<0.01$\\
\hline
\hline
\end{tabular}
\caption{\label{tab:para_z1}
Same as table~\ref{tab:para} but at $z=1$.}
\end{table}
\begin{figure}[tbp]
\centering 
\includegraphics[width=0.85\textwidth=0 380 0 200]{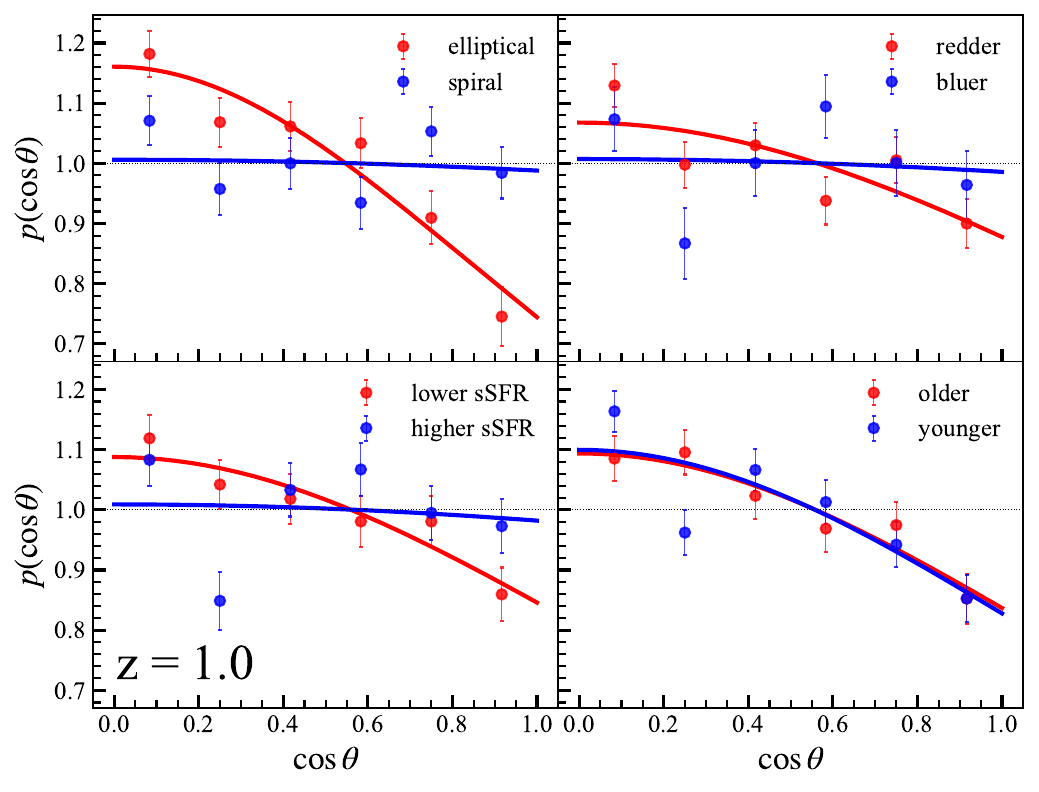}
\caption{\label{fig:pcost_dep_z1} Same as figure~\ref{fig:pcost_dep} but at $z=1$.}
\end{figure}

Given the results presented in sections~\ref{sec:analysis} and~\ref{sec:dep}, we speculate that the expansion of cosmic voids may be at least partially responsible for the transformation 
of void-surface galaxies into elliptical, redder and more quiescent ones. To test this speculation, we investigate the void-surface galaxy perpendicular alignments at two higher redshifts, 
$z=0.5$ and $1$, when the universe contains higher fractions of younger, bluer, and accreting galaxies of spiral types than at $z=0$. 
The second and third rows of table~\ref{tab:void} provides information on the total number, mean effective radius, and mean density contrast of 
the voids identified via the Void-Finder algorithm at $z=0.5$ and $1$, respectively. 
If the expansion of cosmic voids truly plays a crucial role in transforming the morphology, color and sSFR of the galaxies on their surfaces, then a similar 
trend of the property dependence of the void-surface galaxy perpendicular alignments would be found even at higher redshifts, since the higher-$z$ voids as those at $z=0$ 
should have the same effects on the their surface galaxies as those at $z=0$, while causing their perpendicular alignments. If the high-$z$ voids and their surface galaxies do 
not exhibit a similar property-dependence of perpendicular alignments, then it would provide a strong counter-evidence against our speculation, indicating that 
the stronger alignment signals of ellipticals, redder and more quiescent galaxies at $z=0$ must be caused by earlier mitigation of those galaxies to the void-surfaces after they have 
already become stalled and quenched somewhere else.

Re-following the same procedures described in sections~\ref{sec:mor_col} and~\ref{sec:ssfr_age} to the voids at $z=0.5$ and $1$, we dichotomize the void-surface galaxies, 
according to each of the four properties. Applying the same cut-offs of $m_{\star}\ge 10.5$ and $0.9\le r/R_{\rm v}\le 1.1$ as at $z=0$,  we determine $p(\cos\theta)$ and $d_{t}$ 
from each subsample, which are shown in figures~\ref{fig:post_dep_z0.5}-\ref{fig:pcost_dep_z1}  and tables~\ref{tab:para_z0.5}-\ref{tab:para_z1}. 
As can be seen, very similar property dependences of $p(\cos\theta)$ and $d_{t}$ are found at both of the two redshifts. The signals of perpendicular alignments are significant 
only for the case of those void-surface galaxies that have elliptical types, redder colors, and lower sSFR.  Note, however, an apparent difference in the dependence of $p(\cos\theta)$ 
and $d_{t}$ on the stellar formation epochs, $t_{f\star}$ between $z=0.5$ and $z=1$.  The alignment signals seem to be independent of $t_{f\star}$ at $z=1$, which is different 
from the trends found at $z=0$ and $0.5$ when the older void-surface galaxies yield much stronger signals than the younger counterparts. 

This difference between the lower and higher redshifts stems from the fact that the void-surface galaxies observed at $z=1$ have very similar formation epochs. In other words, 
the two samples containing younger and older void-surface galaxies at $z=1$ do not differ much from each in their distribution of $t_{f\star}$, which in turn causes almost no difference 
in $p(\cos\theta)$ and $d_{t}$.
Even though the void-surface galaxies at $z=1$ have quite similar formation epochs, the perpendicular alignments of their shape axes with the directions toward the void centers 
are found to depend strongly on their morphological types, colors and sSFR.  Although this result is consistent with our speculation, it is not sufficient to prove directly 
that the expansion of cosmic voids indeed contribute to stalling and quenching their surface galaxies. A direct evidence would require to trace the merger tree of each void-surface 
galaxy and then to inspect its properties when it arrives at its host void, which we believe is beyond the scope of this paper.

\section{Robustness and feasibility tests of the alignment signals}\label{sec:robust}

\subsection{Dependence on the void-identification scheme}\label{sec:algorithm}

Now that we have detected clear signals of the void-surface galaxy perpendicular alignments and their dependences on the galaxy properties, 
we would like to examine if the signals are robust against the variation of void-identification schemes. Unlike the case of bound structures that 
are in principle defined to satisfy the virial condition,  there is no unique definition of voids nor any established condition for their formation. 
A variety of void-identification algorithms has so far been proposed, each of which has its own merits and downsides, complementing 
one another.  One of the best merits of the Void-Finder algorithm~\cite{HV02} that we have exclusively employed for the current work 
is that since it finds a void as an assembly of overlapped empty spherical shells, it can expedite the process of locating the void-surface galaxies. 
Nevertheless, it suffers from its parameter dependence (i.e., dependence on the values of $l_{c}$ and $d_{3}$ described in section~\ref{sec:procedure}), 
which in turn sensitively depend on particle mass-resolution in simulations and the flux-limit of galaxy surveys in observations. 

To verify that the detected signals of the void-surface galaxy perpendicular alignments are not spurious ones caused by a particular choice of $l_{c}$ and $d_{3}$, 
we employ a parameter-free void-identification scheme, called the VIDE (Void IDentification and Examination), developed by ref.~\cite{vide}.
The VIDE is an improved version of the ZOBOV (ZOnes Bordering On Voidness), originally developed by ref.~\cite{zobov}, that conducts the Voronoi-tessellation-based reconstruction 
of the density fields and determines the most underdense regions as voids through a watershed transform of reconstructed density fields. The improvement was made in the direction 
of enhancing the speed and flexibility of the ZOBOV code to make the VIDE applicable even to the observational surveys of non-regular volumes. 
In the current work, we utilize the publicly available VIDE python code to the TNG 300-1 galaxy catalog to re-identify the voids and repeat the whole computation described in 
sections~\ref{sec:analysis}-\ref{sec:dep} to see whether or not the vide voids also yield similar alignment trends. 
The total number, mean effective radius and mean density contrasts of the voids identified via the VIDE at $z=0$ are found to be $5337$, $5.15 h^{-1}{\rm Mpc}$  and $-0.43$, respectively.

\begin{figure}[tbp]
\centering 
\includegraphics[width=0.85\textwidth=0 380 0 200]{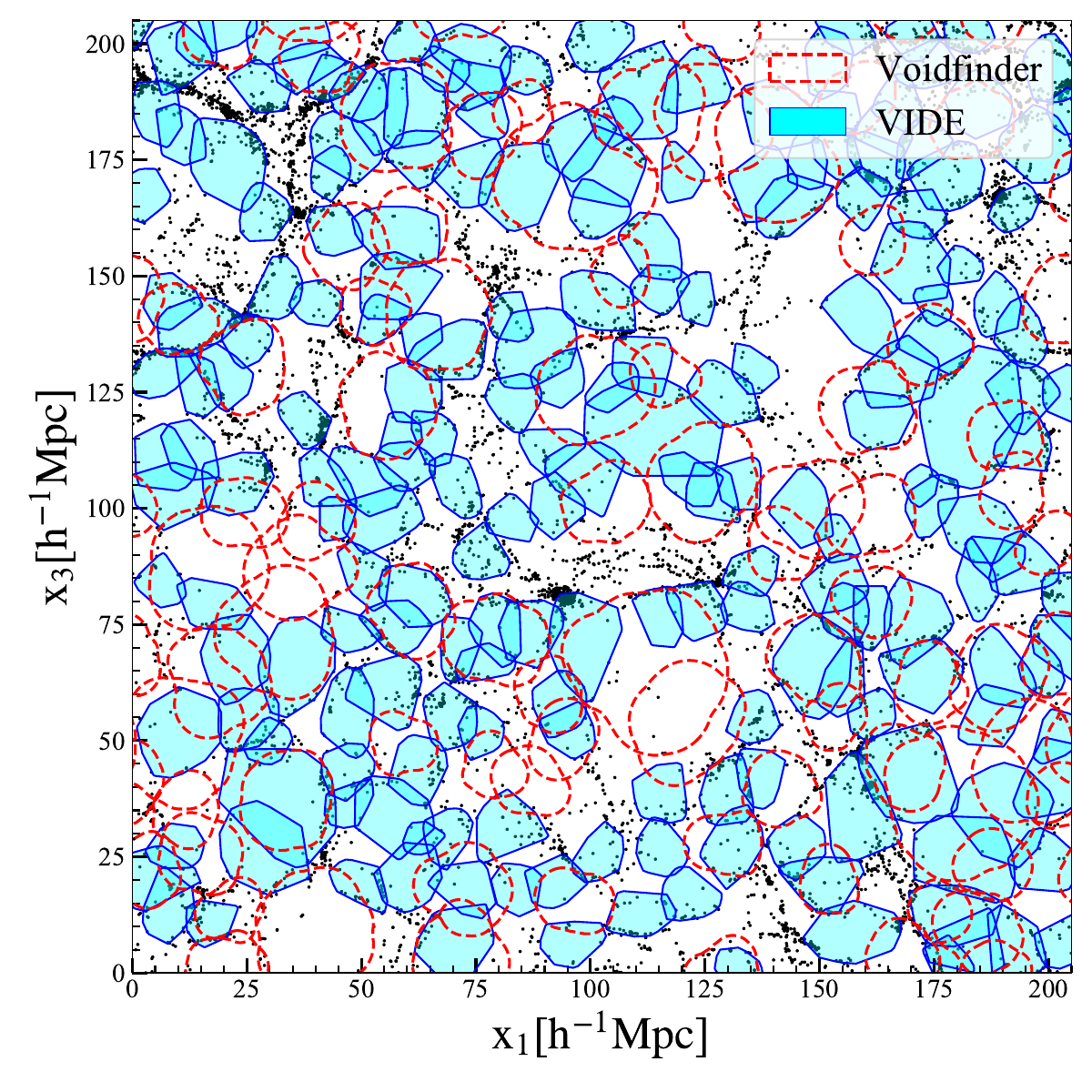}
\caption{\label{fig:vide} Voids identified via the VIDE (sky-blue regions) from the spatial distributions of TNG300-1 galaxies (black dots) compared 
with the Void-Finder voids (regions enclosed by the red dashed curves) in the two-dimensional space projected onto the $\hat{y}$-axis.}
\end{figure}

\begin{figure}[tbp]
\centering 
\includegraphics[width=0.85\textwidth=0 380 0 200]{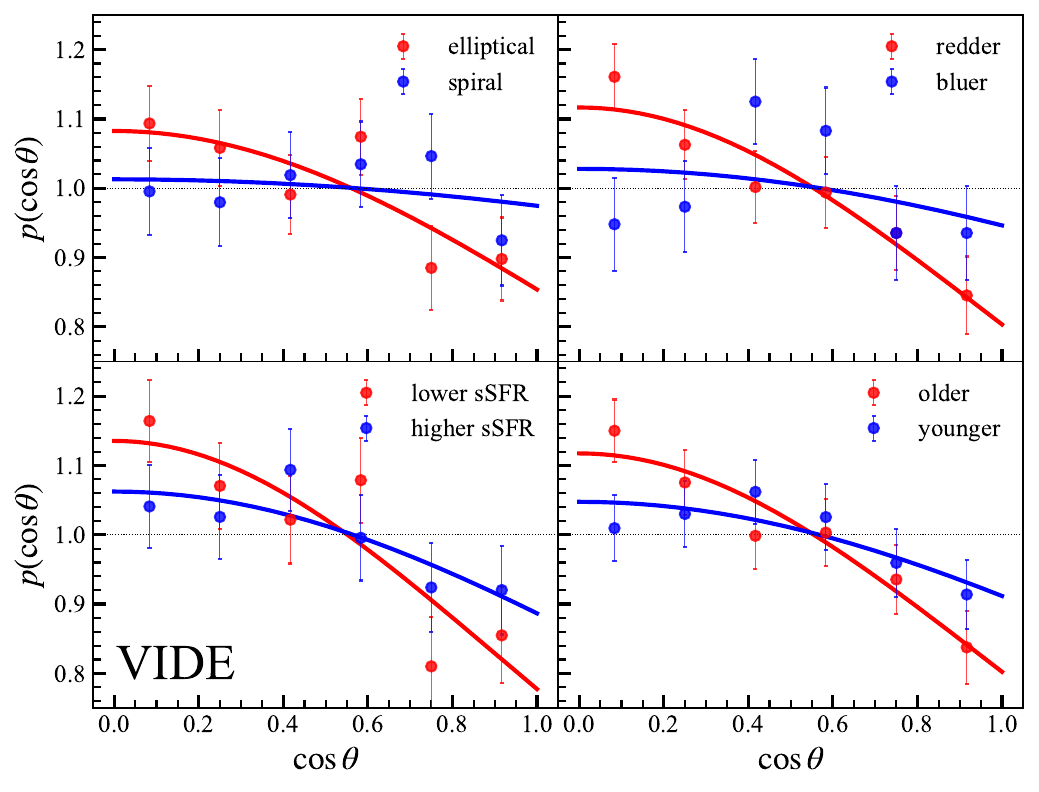}
\caption{\label{fig:pcost_vide} Same as figure~\ref{fig:pcost_dep} but for the case that the voids are identified via the VIDE algorithm~\cite{vide} at $z=0$.}
\end{figure}

Figure~\ref{fig:vide} show the two dimensional projected images of multiple VIDE voids (sky-blue regions within blue curves) and compare them with the Void-Finder voids 
(regions within dashed red curves), from the spatial distributions of the TNG300-1 galaxies (black dots).  The same cut-off of $m_{\star}\ge 10.5$ and $0.9\le r/R_{\rm v}\le 1.1$ 
as for the Void-Finder voids are applied to the VIDE voids to determine $p(\cos\theta)$.
Table~\ref{fig:pcost_vide} and figure~\ref{fig:pcost_vide} show the same entries as table~\ref{fig:pcost} and figure~\ref{fig:pcost_vide}, respectively, but for the case of the VIDE voids.  
The same cut-off of $m_{\star}\ge 10.5$ and $0.9\le r/R_{\rm v}\le 1.1$ as for the Void-Finder case are applied to the VIDE voids to determine $p(\cos\theta)$ and $d_{t}$. 
As can be seen, the VIDE voids exhibit lower signal-to-noise ratio, although they yield quite similar trends of the void-surface galaxy perpendicular alignments and their dependences 
on the four galaxy properties.  

The difference in the signal-to-noise ratios between the Void-Finder and VIDE cases may be related to the difference between the two algorithms in defining the void boundaries. 
The boundary of a void in the former algorithm coincides with the rim of an overlapped empty {\it spherical shells}, while the latter makes no presumption about the geometrical 
shapes of a void region found in the density field constructed on the Voronoi-tessellation cells~\cite{zobov}.  Although this {\it spherical rim} presumption of the Void-Finder algorithm about the 
void boundaries could be regarded as its limitation or downside, it seems to make the key approximation work better for the determination of the void-surface galaxy perpendicular alignments, 
the approximation that the directions from the void-surface galaxies to the void centers are parallel to the major principal axes of the local tidal fields. 
Besides, it is consistent with the finding of the previous work~\cite{LM23} that the spin transitions of spiral galaxies on void exhibit much higher signal-to-noise ratios when the voids 
were identified via the Void-Finder algorithm rather than via the ZOBOV schemes.  The comparison between the results from the Void-Finder and VIDE algorithms implies  that the 
Void-Finder, albeit classical, may be the optimal for the investigation of spin/shape alignments of void-surface galaxies with the directions to void centers.

\begin{table}[tbp]
\centering
\begin{tabular}{cccccccc}
\hline
\hline
\rule{0pt}{4ex}\noindent
${\rm criterion}$ & $N_{\rm ori}$ & $N_{\rm cont}$ & $10^{2} d_{t}$ & $\langle \cos \theta \rangle$ & $p_{\rm 3D}$ \\
\hline
\rule{0pt}{4ex}\noindent
$\kappa < 0.45$ & $30032$ & $13613$ & $5.11 \pm 1.77$ & $0.48 \pm 0.01$ &$0.15$\\
$\kappa \ge 0.45$ & $11273$ & $11272$ & $0.81 \pm 1.33$ & $0.50 \pm 0.01$ &$0.64$\\
\hline
\rule{0pt}{4ex}\noindent
$g-r \ge 0.75$ & $30853$ & $17090$ & $7.15 \pm 1.66$ & $0.47 \pm 0.01$ &$0.05$\\
$g-r < 0.75$ & $10452$ & $10446$ & $1.75 \pm 1.84$ & $0.49 \pm 0.01$ &$0.66$\\
\hline
\rule{0pt}{4ex}\noindent
$\log({\rm sSFR}) < -12.5$ & $29724$ & $11437$ & $8.27 \pm 2.07$ & $0.47 \pm 0.01$ &$0.01$\\
$\log({\rm sSFR}) \ge -12.5$ & $11581$ & $11434$ & $3.87 \pm 1.90$ & $0.48 \pm 0.01$ &$0.41$\\
\hline
\rule{0pt}{4ex}\noindent
$t_{f\star} < 5.43\, {\rm Gyr}$ & $20021$ & $19478$ & $7.20 \pm 1.55$ & $0.47 \pm 0.01$ &$0.02$\\
$t_{f\star} \ge 5.43\,{\rm Gyr}$ & $21261$ & $19478$ & $2.96 \pm 1.45$ & $0.49 \pm 0.01$ &$0.34$\\
\hline
\hline
\end{tabular}
\caption{\label{tab:para_vide}
Same as table~\ref{tab:para} but from the VIDE voids.}
\end{table}

\subsection{Redshift-space effects on the alignment strengths}\label{sec:rsd}
\begin{table}[tbp]
\centering
\begin{tabular}{cccccccc}
\hline
\hline
\rule{0pt}{4ex}\noindent
${\rm criterion}$ & $N_{\rm ori}$ & $N_{\rm cont}$ & $10^{2} d_{t}$ & $\langle \cos \theta \rangle$ & $p_{\rm 3D}$ \\
\hline
\rule{0pt}{4ex}\noindent
$\kappa < 0.45$ & $29997$ & $13564$ & $8.36 \pm 1.32$ & $0.47 \pm 0.00$ &$<0.01$\\
$\kappa \ge 0.45$ & $11290$ & $11289$ & $1.66 \pm 1.38$ & $0.49 \pm 0.01$ &$0.53$\\
\hline
\rule{0pt}{4ex}\noindent
$g-r \ge 0.68$ & $37032$ & $18114$ & $7.20 \pm 1.16$ & $0.47 \pm 0.00$ &$<0.01$\\
$g-r < 0.68$ & $4255$ & $4251$ & $0.51 \pm 1.30$ & $0.50 \pm 0.01$ &$0.98$\\
\hline
\rule{0pt}{4ex}\noindent
$\log({\rm sSFR}) < -10.4$ & $39408$ & $17428$ & $5.62 \pm 1.13$ & $0.48 \pm 0.00$ &$<0.01$\\
$\log({\rm sSFR}) \ge -10.4$ & $1879$ & $1875$ & $1.39 \pm 2.29$ & $0.50 \pm 0.01$ &$0.97$\\
\hline
\rule{0pt}{4ex}\noindent
$t_{f\star} < 5.43\,{\rm Gyr}$ & $19995$ & $19458$ & $10.29 \pm 1.25$ & $0.46 \pm 0.00$ &$<0.01$\\
$t_{f\star} \ge 5.43\,{\rm Gyr}$ & $21269$ & $19458$ & $5.43 \pm 1.11$ & $0.48 \pm 0.00$ &$<0.01$\\
\hline
\hline
\end{tabular}
\caption{\label{tab:para_rsd}
Same as table~\ref{tab:para} but in redshift-space for which the line-of-sight directions are assumed to be parallel to the Cartesian $\hat{x}_{3}$-axis.}
\end{table}

\begin{figure}[tbp]
\centering 
\includegraphics[width=0.85\textwidth=0 380 0 200]{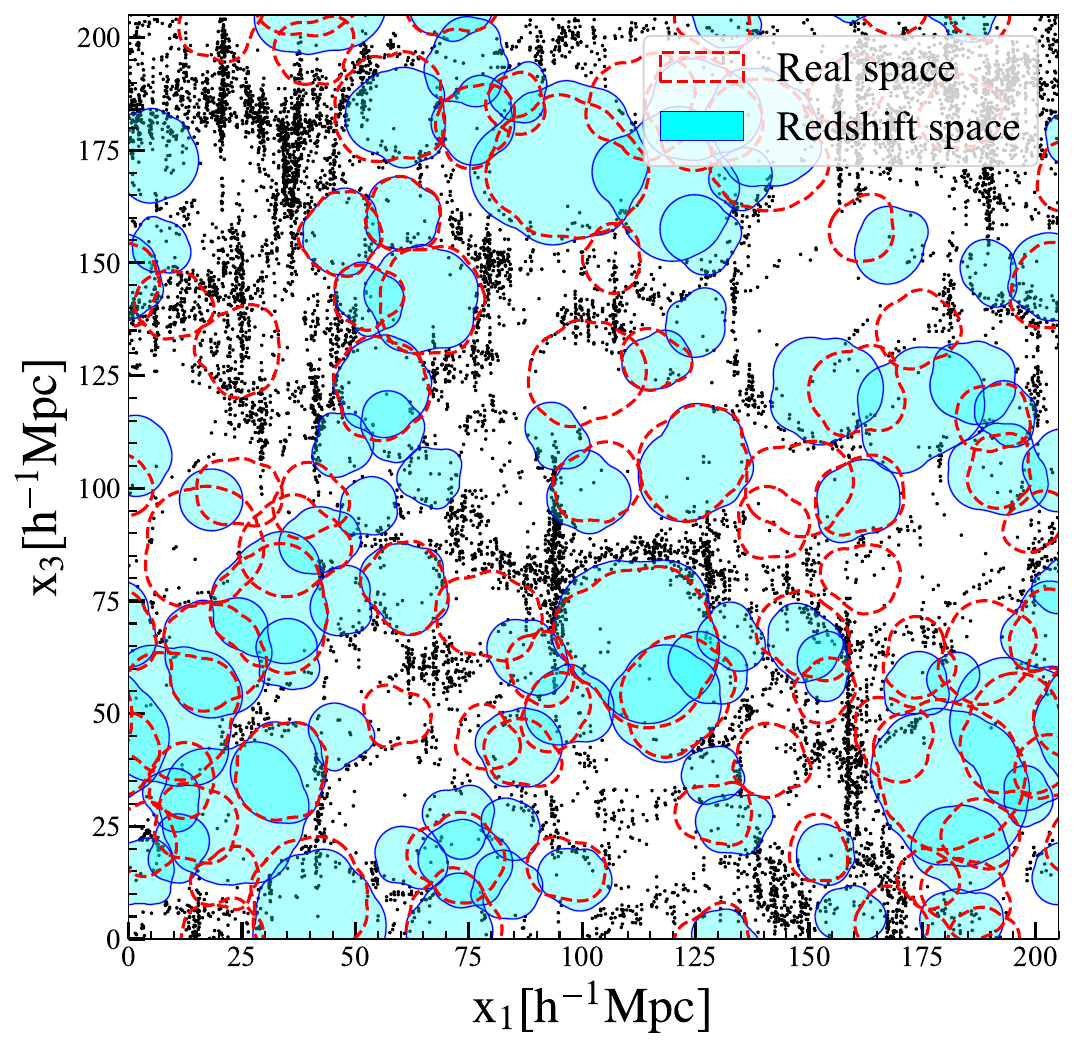}
\caption{\label{fig:rsd} Two-dimensional projected images of the Void-Finder voids (sky-blue regions) identified in the redshift-space galaxy distributions (black dots), 
compared with those in the real-space counterparts (regions enclosed regions) at $z=0$.}
\end{figure}
\begin{figure}[tbp]
\centering 
\includegraphics[width=0.85\textwidth=0 380 0 200]{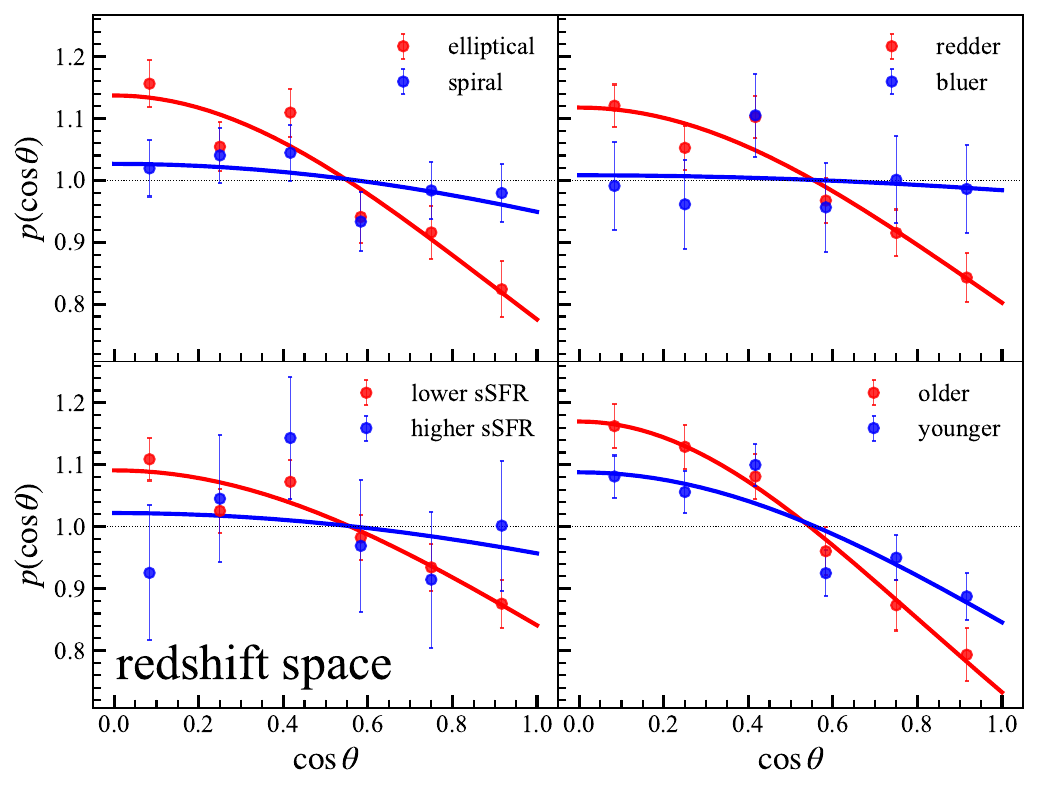}
\caption{\label{fig:pcost_rsd} Same as figure~\ref{fig:pcost_dep} but in redshift-space for which the line-of-sight directions are assumed to be parallel to the Cartesian $\hat{x}_{3}$-axis at $z=0$.}
\end{figure}

To detect a signal of the void-surface galaxy perpendicular alignments from real observations, it must be taken into account that the voids can readily be identified only in redshift space 
and thus that the shapes and boundaries of voids found in redshift-space are quite likely to differ from those in real space.
For the exploration of the redshift-space effects on the void-surface galaxy perpendicular alignments, we create redshift-space galaxy distributions as follows, under the flat-sky approximation 
that the line-of-sights of all galaxies are parallel to the $\hat{\bf x}_{3}$-axis: each real-space galaxy position, $(x_{1},x_{2},x_{3})$ modified into $(x_{1},x_{2},x_{r3})$ with 
$x_{r3}\equiv x_{3} + {\bf v}\cdot\hat{\bf x}_{3}/H_{0}$ where ${\bf v}$ is the peculiar velocity of each galaxy and $H_{0}=100\,h\,{\rm km}\,s^{-1}\,{\rm Mpc}^{-1}$ 
with dimensionless Hubble constant $h$.

Repeating the whole analysis presented in sections~\ref{sec:analysis}-\ref{sec:dep} but with the redshift-space voids and their surface galaxies, we determine $p(\cos\theta)$ and 
$d_{t}$.  Figure~\ref{fig:rsd} shows the real-space voids (sky-blue regions within sold blue curves) from the spatial distributions of redshift-space galaxies (black dots) and compares 
them with the real-space counterparts (regions within red dashed curves).  
Figure~\ref{fig:pcost_rsd} displays $p(\cos\theta)$ obtained from the redshift-space void-surface galaxies and its dependence on the four galaxy properties. 
The best-fit values of $d_{t}$ and $p$-values from the KS test are listed in table~\ref{tab:para_rsd}. 
As can be seen, the redshift-space effects on the strengths of the void-surface galaxy perpendicular alignments seem to be not so strong. 
Although the alignments become a weaker, very similar trends of $p(\cos\theta)$ are found for each case.  

\subsection{Projection effects on the alignment strengths}\label{sec:2d}
\begin{figure}[tbp]
\centering 
\includegraphics[width=0.85\textwidth=0 380 0 200]{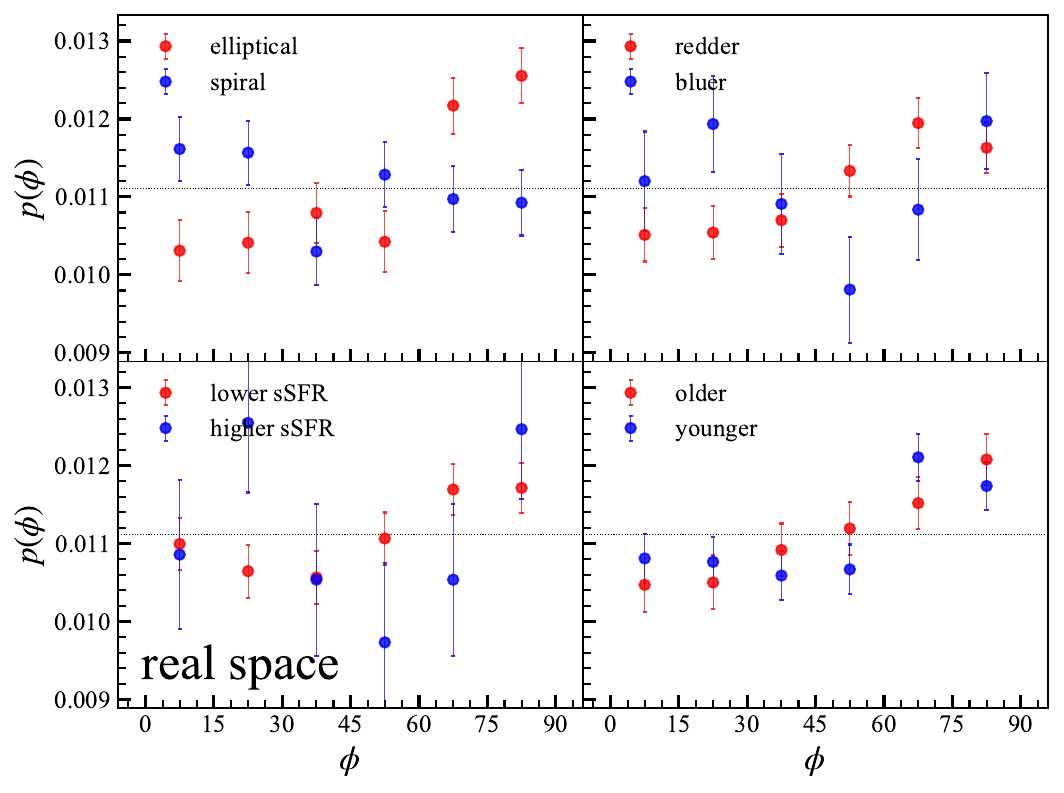}
\caption{\label{fig:pcost_2d} Dependences of $p(\phi)$ with Poisson errors on the four galaxy properties measured in the two-dimensional projected plane 
normal to the Cartesian $\hat{x}_{3}$-axis at $z=0$.}
\end{figure}
\begin{figure}[tbp]
\centering 
\includegraphics[width=0.85\textwidth=0 380 0 200]{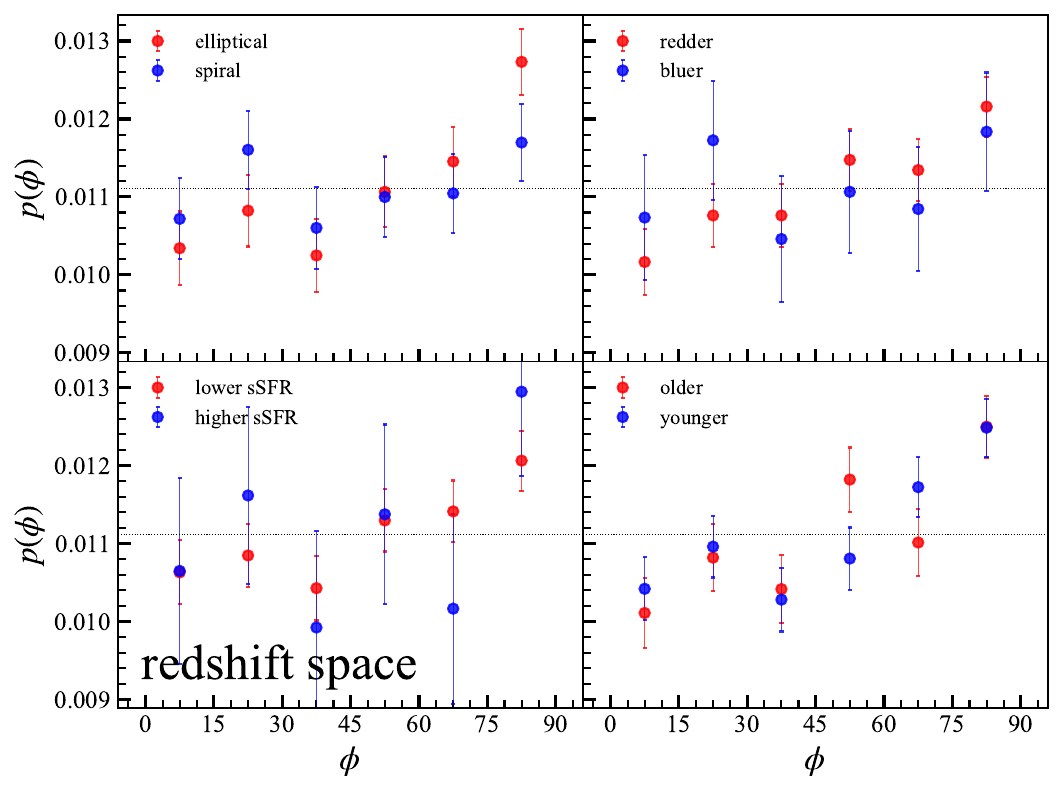}
\caption{\label{fig:pcost_2d_rsd} Same as figure~\ref{fig:pcost_2d} but in the redshift-space.}
\end{figure}
In practice,  what can be readily observable is the orientations not of three dimensional but of two-dimensional (2D) projected shapes of the giant galaxies 
in the sky-plane normal to the line of sight directions. 
As a feasibility test, we project both of ${\bf u}$ and the major principal axes of $(T_{ij})$ parallel to the directions toward the void centers onto the plane normal to the line-of-sight directions, 
$\hat{\bf x}_{3}$-axis. Then, we determine the probability density distributions, $p(\phi)$, where $\phi$ is the angle between the two projected axes in the Cartesian $\hat{x}$-$\hat{y}$ plane, 
and investigate the dependence of $p(\phi)$ on the galaxy properties, performing the KS test, as done in sections~\ref{sec:analysis}-\ref{sec:dep}.  
Note that for the 2D case, we determine $p(\phi)$ rather than $p(\cos\phi)$ since it is $p(\phi)$ that would become uniform if there is no alignment. 
The uncertainty in the determination of $p(\phi)$ is computed as $1\sigma$ Poisson error as $\sigma = 1/(90\sqrt{n_{g}-1})$ since $p(\phi)$ becomes $1/90$ if there is no alignment tendency. 

Each panel of figures~\ref{fig:pcost_2d}-\ref{fig:pcost_2d_rsd} plots $p(\phi)$ in the range of $[0,\ 90^{\circ}]$ with Poisson errors from two $m_{\star}$-controlled samples dichotomized 
by each of the four galaxy properties in real and redshift spaces, respectively.  
As can be seen and as expected, the 2D projection has an effect of reducing the overall strengths of the void-surface galaxy perpendicular alignments and enlarging the errors as well. 
In real space, a significant signal of void-surface galaxy perpendicular 2D alignments is detected only from the elliptical galaxies, while in redshift space, no significant signal is found even 
from the elliptical galaxies. This result implies that for a detection of significant signals of the void-surface galaxy perpendicular alignments and their property dependence in practice, 
what is required is to reduce the associated errors by statistically reconstructing the real-space voids and/or by stacking them over a wide range of redshifts~\cite{stack_voids}.
 
 \section{Summary and discussion}\label{sec:con}

Identifying the voids and finding the galaxies on their surfaces by applying the Void-Finder algorithm~\cite{HV02} to the TNG 300-1 simulation data~\cite{tng1,tng2,tng3,tng4,tng5,tng6}, 
we have numerically determined the probability density distributions of the cosines of the angles, $p(\cos\theta)$, between the shape axes of the giant void-surfaces galaxies  
and the directions toward the void centers at $z=0,0.5$ and $1$.  We have detected a clear signal of perpendicular alignments between them and found that the alignment strengths depend 
on the galaxy morphologies, colors, and sSFR.   We have also shown that the behavior of $p(\cos\theta)$ is well described by the one-parameter model, eq.~(\ref{eqn:pcost}), 
derived by ref.~\cite{lee19} under the assumption that the shape axes of giant galaxies develop a tendency to lie in the plane normal to the major principal axis of the local tidal field. 

The following summarizes the key results of the current work.  
\begin{itemize}
\item
At each redshift, significant signals of void-surface galaxy perpendicular alignments are found from those giant galaxies with stellar masses 
$m_{\star}\equiv\log[M_{\star}/(h^{-1}\,M_{\odot})]\ge 10.5$ located on void surfaces in the radial distance ranges of $0.9\le r/R_{\rm v}\le 1.1$ with radii of empty spheres $R_{\rm v}$ 
that consist of a hosting void (see figures~\ref{fig:mcost}-\ref{fig:pcost}). 
\item
Between the two $m_{\star}$-controlled samples of ellipticals and spirals on void surfaces at $z=0$, only the former yields strong signals of 
void-surface galaxy perpendicular alignments, while no signal is found from the latter. The mean alignment strength seems to almost monotonically increase
as the fraction of the rotational kinetic energy, $\kappa$, decreases (see figures~\ref{fig:pcost_dep} and~\ref{fig:pcost_var}). 
\item
When the void-surface galaxies is dichotomized into $m_{\star}$-controlled subsamples according their $g$-$r$ colors, only the redder galaxies yield significant 
signals of perpendicular alignments, while no signal is found from the bluer ones.
As the galaxies become redder and redder, the void-galaxy perpendicular alignments become stronger and stronger (see figures~\ref{fig:pcost_dep} and~\ref{fig:pcost_var}). 
\item
Between the two $m_{\star}$-controlled samples of the void-surface galaxies with lower and higher sSFR, a significantly stronger signal of 
void-surface galaxy perpendicular alignments is detected from the former.  The void-surface galaxy perpendicular alignments show a consistent trend of becoming stronger
as the sSFR of void-surface galaxies diminishes (see figures~\ref{fig:pcost_dep} and~\ref{fig:pcost_var}). 
\item
The void-surface galaxy perpendicular alignments exhibit only weak dependence on the stellar age, $t_{f\star}$, of the void-surface galaxies. 
No threshold of $t_{f\star}$ is able to create a sample that yield negligibly low signal of the void-surface galaxy perpendicular alignments. 
\item
The analytic formula devised by ref.~\cite{lee19} has been proven to match the numerical results of $p(\cos\theta)$ at all of the three redshifts, and its correlation parameter has 
turned out to efficiently quantify the strengths of void-galaxy perpendicular alignments and their variations with galaxy properties (see table~\ref{tab:para}). 
\item
Using the parameter-free VIDE algorithm~\cite{vide} that is quite different from the parameter-dependent Void-Finder, we have tested the robustness of our results. 
It has been found that although the VIDE voids yield weaker signals of the void-surface galaxy perpendicular alignments, the overall trend of the property dependence of $p(\cos\theta)$ 
is very similar to that from the Void-Finder voids. The difference in the alignment strengths between the Void-Finder and VIDE voids has been ascribed to the difference in the definition 
of void boundaries between the two. 
\item
Taking into account the redshift-space and projection effects, we have tested the feasibility of detecting the void-surface galaxy perpendicular alignments and their property dependences.  
While little difference is found between the redshift and real 3D spaces in the strengths of void-surface galaxy perpendicular alignments and their dependences on the galaxy 
properties (see figure~\ref{fig:pcost_rsd}), the 2D projection effects have turned out to considerably weaken the alignment signals in the redshift space 
(see figures~\ref{fig:pcost_2d}-\ref{fig:pcost_2d_rsd}). This result has implied that a large sample of stacked voids and void-surface galaxies should be required to reduce the statistical 
errors for a detection of significant signals in redshift 2D space. 
\end{itemize}

The key results of the current analysis hints that the cosmic voids may contribute to the stalling and quenching of the galaxies located on their surfaces. 
The formation of a void and its rapid expansion along the major principal axis of the large-scale tidal field tightly compresses the neighboring matter, 
leading the adjacent galaxies to develop strong perpendicular shape alignments with the directions to its center.  This matter compression caused by 
the void expansion may also generates particle flows orthogonal to its surface, preventing the gas particles from radial falling and accreting onto the void-surface galaxies, which could 
effectively stall and quench them.  Henceforth, if a galaxy adjacent to a void is exposed to the effect of a rapid void expansion for a longer period of time, then its shape would develop 
a stronger perpendicular alignment with the direction to the void center. At the same time, it would become more quickly redder and quiescent, and its rotational motion would convert 
more rapidly into random counterparts. Whereas, the galaxies that have recently migrated to the surface of a given void long after its formation would have exhibit no significant signal of 
void-surface galaxy perpendicular alignments. 

In this speculative picture,  the abundance of elliptical void-surface galaxies with strong void-surface galaxy perpendicular alignments might depend on the dynamical nature of 
dark energy. According to the latest results from the Dark Energy Spectroscopic Instrument (DESI)~\cite{desi}, 
the cosmic acceleration at the present epoch is likely to be caused not by the completely inert cosmological constant~\cite{lambda}, but by a dynamical dark anergy whose equation of state 
evolves with time~\cite{quint99,quint03,quint08}. Since the evolution of dark energy equation of state would change the formation and expansion rates of giant cosmic voids.
it could be envisioned that in some dynamical DE models where the cosmic voids form early and expand faster in the past~\cite{phantom}, there would be a higher number fraction of the 
elliptical galaxies on void surfaces that exhibit stronger void-surface galaxy perpendicular alignments.  This mechanism, i.e., the earlier formation of cosmic voids due to dynamical 
DE and its strong effects of quenching and stalling the high-$z$ galaxies, might be able to resolve another long-standing puzzle that challenges the standard galaxy formation paradigm, 
the presence of massive high-$z$ quenched galaxies, even in gas-rich environments~\cite{highz_quench1,highz_quench2}.

Yet, the current work provides no direct evidence for our speculations. 
One practical way to test this speculation will be to numerically examine if the void-surface galaxies are more promptly quenched compared with the galaxies of similar masses embedded 
in filaments and knots. If it turns out to be the case, then it would prove that the matter compression caused by the void expansion is indeed an efficient extra mechanism of stalling and 
quenching the galaxies on void surfaces. 
A more direct test could be performed by numerically tracking down the assembly history (or merger tree) of each void-surface galaxy and then classifying them into long-term residents 
and recent arrivals.  If the signals of void-surface perpendicular alignments turn out to differ between them, it would directly verify that the quenching and stalling actually requires the extended 
exposure, the key tenet of our scenario.  
Regarding the proposal that the void-surface galaxy perpendicular alignments could be a complementary probe of the nature of DE, it would be desirable to explicitly investigate 
if and how the alignment signals vary with the equation state of dark energy by analyzing the data from hydrodynamical simulations that run for dynamical dark energy models.
Our future work will be in the directions of conducting these follow-up tasks as well as of exploring the void-surface galaxy perpendicular alignments from real observations. 

\acknowledgments

The IllustrisTNG simulations were undertaken with compute time awarded by the Gauss Centre for Supercomputing (GCS) 
under GCS Large-Scale Projects GCS-ILLU and GCS-DWAR on the GCS share of the supercomputer Hazel Hen at the High 
Performance Computing Center Stuttgart (HLRS), as well as on the machines of the Max Planck Computing and Data Facility 
(MPCDF) in Garching, Germany.  
JL acknowledges the supports by Basic Science Research Program through the National Research Foundation (NRF) 
of Korea funded by the Ministry of Education (RS-2025-00512997).

\end{document}